\documentstyle[amsmath,psfig]{article}

\newcommand{\bold}[1]{\mbox{\boldmath $#1$}}
\newcommand{\XOR}{\oplus}
\newcommand{\OR}{\vee}
\newcommand{\AND}{\wedge}
\newcommand{\go}{\rightarrow}
\newcommand{\tru}[1]{\left[\!\left[#1\right]\!\right]}
\newcommand{\mcal}[1]{{\bold{#1}}}
\begin{document}
\title{CELLULAR AUTOMATA}
\author{FRANCO BAGNOLI\\	
   Dipartimento di Matematica Applicata\\ 
	Universit\`a di Firenze, via S. Marta, 3\\
	I-50139 Firenze Italy\\
	e-mail: bagnoli@dma.unifi.it\\
\\ \small
This paper will appear in \\
\small {\bf Dynamical Modelling in Biotechnologies},\\
\small  F. Bagnoli, P. Li\'o and S. Ruffo, editors (World Scientific, Singapore, 
1998) 
}
\maketitle

\section{Introduction}
In this lecture I shall  present a class of  mathematical tools for
modeling phenomena that can be described in terms of elementary
interacting objects.  The goal is to make the macroscopic
behavior arise from individual dynamics. I shall denote these individuals
with the term automaton, in order to emphasize the main ingredients of the
schematization: parallelism and locality. In my opinion, a \textit{good}
microscopic model is based on a rule that can be executed in parallel by
several automata, each of which has information only on what happens in its
vicinity (that can extend arbitrarily). 
In the following I shall use the word automaton either to refer to a single
machine or to the whole set of machines sharing the same evolution rule.
 
These are completely discrete models: the time increases by finite steps,
the space is represented by a regular lattice, and also the possible
states of the automaton (the space variables) can assume one out of a
finite set of values. The reason for this choice is the conceptual
simplicity of the description. A single real number requires an infinity
of information to be completely specified, and since we do not have any
automatic tool (i.e. computer) that efficiently manipulates real numbers,
we have to resort to approximations.  On the other hand, these discrete
models can be exactly implemented on a computer, which is a rather simple
object (being made by humans). Since the vast majority of present
computers are serial ones (only one instruction can be executed at a
time), the parallel nature of the model has to be simulated.

The goal of a simple microscopic description does not imply that one cannot use
real numbers in the actual calculations, like for instance in mean field
approximations. 

The class of phenomena that can be described with automata models is very
large. There are real particle-like objects, such as atoms or
molecules (from a classical point of view), that can be used to model the
behavior of a gas. But one can build up models in which the automata
represents bacteria in a culture or cells in a human body, or patches of
ground in a forest.

In reality there are two classes of automata, one in which the automata
can wander in space, like molecules of a gas, and another one in which the
automata are stick to the cell of a lattice. I shall call the first type
\textit{molecular automata}, and the second \textit{cellular
automata}.~\footnote{See also the contribution by N. Boccara, this volume.}

Probably the first type is the most intuitive one, since it resembles actual
robots that sniff around and move. Each class has its own advantages, in term
of simplicity of the description. A molecular automaton has information about
its identity and its position. It can be used to model an animal, its state
representing for instance the sex and the age. It is quite easy to write down
a rule to make it respond to external stimuli. However, one runs into troubles
when tries to associate a finite spatial 
dimension to this automaton. Let us suppose
that the automaton occupies a cell on the lattice that represents the space.
The evolution rule has to decide what happens if more than one automaton try
to occupy the same cell. This is very hard to do with a true parallel, local
dynamics, and can involve a  negotiation which slows down the simulation.
Clearly, a possible solution is to adopt a serial point of view: choose one of
the automata and let it move. This is the approach of the computations based
on molecular dynamics, where one tries to study the behavior of ensembles of
objects following Newton's equations, or in Monte Carlo calculations. This
serial approach is justified when time is continuous, and the discretization
is just a computational tool, since for a finite number of objects the
probability of having two moves at the same instant is vanishing, but indeed
it is not a very elegant solution. Thus, molecular automata are good for
point-like, non-excluding particles.  

The other solution is to identify the automata with a point in space: on each
cell of the lattice there is a processor that can communicate with the
neighboring ones. If we say that a state represents the empty cell and another
the presence of a bacterium, we associate to it a well defined portion of
space. From this point of view, the bacterium is nothing but a  property of
space. The usual name for this kind of  models is \textit{cellular automata}. 

People from physics will realize that cellular automata correspond to a
field-like point of view, while molecular automata correspond to a particle
point of view. In the last example above, only one particle can sit at a
certain location (cell) at a certain time. Thus, we described a Fermion field.
One can allow also an arbitrary number of particles to share the same
position. This is a Boson field, in which particles loose their
individuality.  Following our analogy with elementary particles, we could say
that molecular automata correspond to classical distinguishable particles. 

The Boson field represents also the link between molecular and cellular
automata. Indeed, if we relax the need of identifying each particle, and if we
allow them to share their state (i.e. their identity and their position), then
the two kinds of automata coincide. This suggests also an efficient way to
simulate a molecular automata. Let us assume that every particle follows a
probabilistic rule, i.e. it can choose among several possibilities with
corresponding probabilities. Consider the case in which there are several
identical particles at a certain position. Instead of computing the fate of
each particle, we can calculate the number of identical particles that will
follow a certain choice. If the number of identical particles at a certain
position is large, this approach will speed up very much the simulation.

In the following I shall concentrate on cellular automata. They have been
introduced in the forties by John von Neumann~\cite{Neumann}, a mathematician
that was also working on the very first computers. Indeed, cellular automata
represent a paradigm for parallel computation, but von Neumann was rather
studying the logical basis of life. The idea of the genetic code was just
arising at that time, so we have to take into consideration  the cultural
period. 

From a mathematical point of view, he realized that the reproduction
process implies that the organism has to include a description of itself,
that we now know to be the genetic code. On the other hand, the
chemistry of real world is too complex, and also the mechanics of a robot is
not simply formalized. The solution was to drastically simplify the world,
reducing it to a two-dimensional lattice. The result of these studies was a
cellular automaton with the power of a general-purpose computer (a Turing
machine), and able to read the information to reproduce itself. This solution
was quite complex (each cell could assume one out of 26 states), but now we
have several simplified versions. One of these is notably based on the Game of
Life, a cellular automaton described in Section~\ref{section:Life}. For a
review of these topics, see~Sigmund (1993)~\cite{Sigmund}. 

In spite of their mathematical interests, cellular automata has been quiescent
for nearly 30 years, until the introduction of the John Conway's Game of Life
in the columns of Scientific American~\cite{Life}, around 1970. The Game of
Life is a two-dimensional cellular automaton in which the
 cells can assume only two
states: 0 for dead and 1 for live. Looking at the evolution of this cellular
automaton an the display of a fast computer is quite astonishing. There are
fermenting zones of space, and Life propagates itself by animal-like
structures. From a prosaic point of view, there are interesting mathematical
questions to be answered, as suggested in Section~\ref{section:Life}.

Finally, cellular automata exited the world of mathematicians around 1984, when
the journal Physica dedicated a whole issue~\cite{Physica10D} to this topic. A
good review of the first application  of cellular automata can also be found
in Wolfram's collection of articles.~\cite{Wolfram}

In the following section I shall consider the mathematical framework of
cellular automata, after which I shall review some applications of the
concept. 

\section{Cellular Automata}

Before going on we need some definitions. I shall denote the spatial index
 $i$ and the temporal index  $t$. Although we can have automata in a
space of arbitrary dimension $d$, it is much simpler for the notations to
consider a one-dimensional space. The state of a cell $\sigma_i^t$ at position
$i$ and at time $t$ can assume a finite number of states. Again for simplicity
I consider only Boolean automata: $\sigma_i^t \in \{0,1\}$. Assume $N$ as the
spatial dimension of the lattice. The state of the lattice can be read as a
base-two number with $N$ digits. Let me denote it with $\bold\sigma^t =
(\sigma_1^t, \dots, \sigma_N^t)$. Clearly $\bold\sigma \in \{0,
2^N-1\}$.~\footnote{In this contribution I shall try to keep notation
constant. I shall denote vectors, matrices and vectorial operators
by bold symbols, including some Boolean functions of an argument that can
take only a finite number of values (like a Boolean string), when referred
as a whole.}

\begin{figure}[t]
\centerline{\psfig{figure=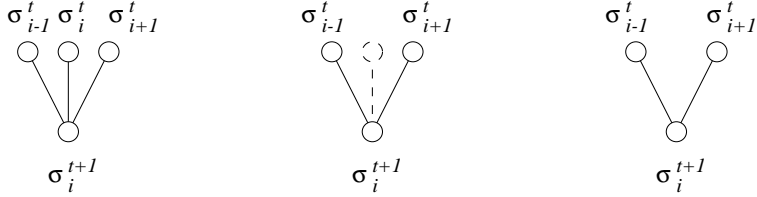,width=10cm,angle=270}}
\caption{The $k=1$ and $k=1/2$ neighborhoods in $d=1$.
\label{neighborhoods}
}
\end{figure}

The evolution rule $\bold\sigma^{t+1} = \bold F (\bold\sigma^t)$ can in
general be written in terms of a local rule 
\begin{equation}
\sigma_i^{t+1} =
f(\sigma_{i-k}^t, \dots, \sigma_i^t, \dots, \sigma_{i+k}^t), 
\end{equation}
where $k$ is the
range of the interactions and the boundary conditions are usually periodic.
The rule $f$ is applied in parallel to all cells. The state of a cell at time
$t+1$ depends on the state of $2k+1$ cells at time $t$ which constitute its
neighborhood. We consider here the simplest neighborhoods: the $k=1$
neighborhood that in $d=1$ consists of three cells and the $k=1/2$
neighborhood, that in $d=1$ consists of two cells and can be considered
equivalent to the previous neighborhood without the central cell. A schematic
diagram of these neighborhoods is reported in Fig.~\ref{neighborhoods}.

\subsection{Deterministic Automata}

The Game of Life and von Neumann's automaton are deterministic ones, i.e.
once given the initial state the fate is in principle known, even though it
can take a lot of effort.

A compact way of specifying the evolution rule for $k=1$, $d=1$ cellular
automata has been introduced by S. Wolfram~\cite{Wolfram}. It consists in
reading all possible configuration of the neighborhood $(\sigma_{i-1}^t,
\sigma_i^t, \sigma_{i+1}^y)$ as a base-two number $n$, and summing up $w_n =
2^n$ multiplied by $\sigma_i^{t+1}$, as shown in Table~\ref{table:r22} for the rule
22.

\begin{table}[t]
\caption{\label{table:r22} Example of Wolfram's code for rule 22.}
\begin {center}
\begin{tabular}{ccc|l|l|c}
\hline
$\sigma_{i-1}^t$ & $\sigma_i^t$ & $\sigma_{i+1}^t$ & 
   $n$ & $w_n$ & $x_i^{t+1}$ \\ 
\hline
0 & 0 & \multicolumn{1}{c|}{0} & 0 & \multicolumn{1}{|r|}{1} & 0 \\ 
0 & 0 & \multicolumn{1}{c|}{1} & 1 & \multicolumn{1}{|r|}{2} & 1 \\ 
0 & 1 & \multicolumn{1}{c|}{0} & 2 & \multicolumn{1}{|r|}{4} & 1 \\ 
0 & 1 & \multicolumn{1}{c|}{1} & 3 & \multicolumn{1}{|r|}{8} & 0 \\ 
1 & 0 & \multicolumn{1}{c|}{0} & 4 & \multicolumn{1}{|r|}{16} & 1 \\ 
1 & 0 & \multicolumn{1}{c|}{1} & 5 & \multicolumn{1}{|r|}{32} & 0 \\ 
1 & 1 & \multicolumn{1}{c|}{0} & 6 & \multicolumn{1}{|r|}{64} & 0 \\ 
1 & 1 & \multicolumn{1}{c|}{1} & 7 & \multicolumn{1}{|r|}{128} & 0 \\ 
\hline
\multicolumn{5}{r}{total} & 22
\end{tabular}
\end{center}
\end{table}

This notation corresponds to the specification of the look-up table for
the Boolean function that constitutes the evolution rule. For an efficient
implementation of a cellular automata one should exploit the fact that all
the bits in a computer word are evaluated in parallel. This allows a
certain degree of parallelism also on a serial computer. This approach is
sometimes called \textit{multi-spin coding} (see
Section~\ref{section:Numerical}). In the following I shall use the symbols
$\XOR$, $\AND$ and $\OR$ for the common Boolean operations e\texttt{X}clusive
\texttt{OR} (\texttt{XOR}), \texttt{AND} and \texttt{OR}. 
The negation of a Boolean variable will be indicated
by a line over the variable. The {\texttt AND} operation will be denoted
often as a multiplication (which has the same effect for Boolean
variables).

Let me introduce some terms that will be used in the following. If a Boolean
function $f$ of $n$ variables $a_1, \dots, a_n$ is completely symmetric with
respect to a permutation of the variables, than it depends only on the value
of the sum $\sum_i a_i$ of these variables, and it is called 
\textit{totalistic}. If
the function is symmetric with respect to a permutation of the variables that
correspond to the values of the cells in the neighborhood, but not to the
previous value of the cell, than the function depends separately on the sum of
the \textit{outer} variables and on the previous value of the cells itself and
the automaton is called \textit{outer totalistic}.

Deterministic cellular automata are discrete dynamical systems. Given a certain
state $\bold\sigma^t$ at a certain time, the state at time $t+1$ is perfectly
determined. The ordered collection of states $(\bold\sigma^0, \dots,
\bold\sigma^t, \dots)$ represents the trajectory of the system.  Since for
finite lattices the number of possible states is also finite, only limit cycles
are possible. There can be several possibilities: only one cycle, several small
cycles, one big and several small, etc, where big and small refer to the length
of the cycle. Moreover, one can consider the basin of attraction of a
cycle, i.e. the number of states that will eventually end on it.

All these dynamical quantities will change with the size of the lattice. A
physicist or a mathematician is generally interested to the asymptotic behavior
of the model, but there may be a well defined size more interesting than
the infinite-size limit. A table with the cycle properties of the simplest
cellular automata can be found at the end of Wolfram's
collection.~\cite{Wolfram}

The study of the limit cycles is clearly limited to very small lattices,
especially in dimension greater than one. To extend the investigation to
larger lattices, one has to resort to statistical tools, like for instance the
entropy of the substrings (patches) of a certain size. 

\begin{figure}[t]
\centerline{\psfig{figure=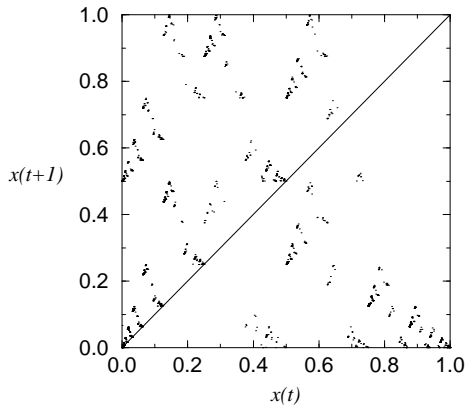,height=6cm}}
\caption{The return map of rule 22.
\label{figure:map22}
}
\end{figure}

If a Boolean string $\bold{a}$, of length $n$ 
appears with probability $p(\bold{a})$ (there are $2^n$ such strings),
 then the normalized $n$-entropy $S_n$ is defined as
\begin{equation}
S_n = -\dfrac{1}{n \ln(2)} \sum_{\bold{a}} p(\bold{a}) \ln (p(\bold{a})).
\end{equation}
$S_n$ ranges from 1, if all strings appear with the same probability
$p=1/2^n$,
to 0 if only one string  appears.

One is generally interested in the scaling of $S_n$ with $n$. 
For an example of
the application of this method, see Grassberger's papers on rule
22.~\cite{Grassberger}

If one reads the configuration $\bold\sigma^t=01001001110101\dots$ as the
decimal digit of the fractional part of a base-two number $x(t)$ (dyadic
representation), i.e.
\begin{equation}
	x(t) = 0.01001001110101\dots
\end{equation}
for an infinite lattice one has a correspondence between the points in the
unit interval and the configurations of the automaton. One has thus a complete
correspondence between automata and maps $x(t+1)=f(x(t))$ of the unit
interval. The map $f$ is generally very structured 
(see Fig.~\ref{figure:map22}). This
correspondence helps to introduce the tools used in the study of dynamical
systems.

\begin{figure}[t]
\centerline{\psfig{figure=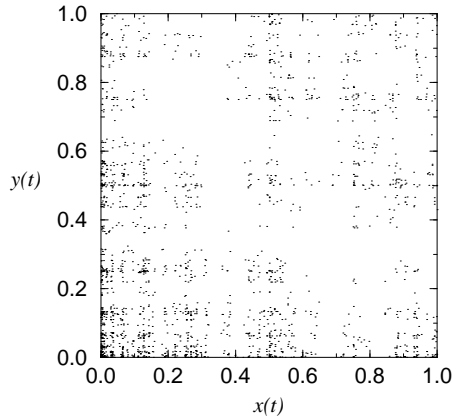,height=6cm}}
\caption{The attractor of rule 22.
\label{figure:attractor22}
}
\end{figure}

Another possibility is the plot of the attractor and the measure of its
fractal dimension. This can be performed dividing the configuration in two
parts, and reading the left (right) part as a decimal number $x(t)$
($y(t)$). The portrait of the attractor for rule 22 is given in 
Fig.~\ref{figure:attractor22}.

One can try to extend to these dynamical systems the concept of chaotic
trajectories. In the language of dynamical systems a good indicator of
chaoticity is the positivity of the maximal Lyapunov exponent, which measures
the dependence of the trajectory with respect
to a small change in the initial position.
This concept can be extended to cellular automata in two ways. The first
solution is to exploit the correspondence between configurations and point in
the unit interval. At the very end, this reduces to the study of the dynamical
properties of the map $f$. This solution is not very elegant, since it looses
the democratic point of view of the system (each cell has the same
importance). 

\begin{figure}[t]
\centerline{\psfig{figure=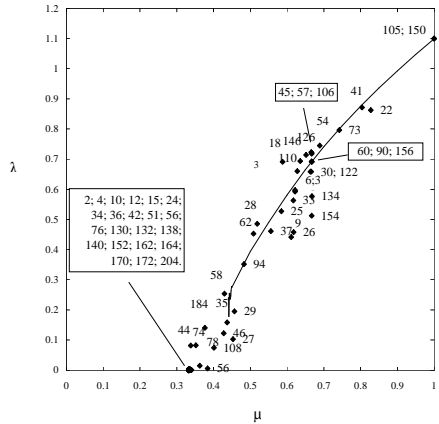,height=6cm}}
\caption{Lyapunov exponent for $k=1$ cellular automata. The symbol $\mu$
denotes the instantaneous diverging rate of trajectories, and the solid line
the results of a random matrix approximation. For further details, see
Bagnoli {\it et al.}\ (1992).~\protect\cite{Bagnoli:Lyapunov}
\label{figure:CALyapunov}
}
\end{figure}

Another possibility is to look at the state of the system as a
point in very high dimensional Boolean space. Here the smallest perturbation
is a change in just one cell, and this damage can propagate at most linearly
for locally interacting automata. However, one can measure the instantaneous
spreading of the damage (i.e. the spreading in one time step), and from here
it is possible to calculate a Lyapunov exponent~\cite{Bagnoli:Lyapunov}. The
automata that exhibit \textit{disordered} patterns have indeed a positive
exponent for almost all starting points (trajectories), while \textit{simple}
automata can have Lyapunov exponent ranging from $-\infty$ to positive values.
As for deterministic dynamical systems one interprets the trajectories with
negative Lyapunov exponents as stable ones, and those with a positive exponent
as unstable ones. In the simulation of an usual dynamical system, rounding
effects on the state variables lead to the disappearance of the most unstable
trajectories, so that the dependence of Lyapunov exponents on the trajectories
looks quite smooth. We can recover this behavior with our discrete dynamical
systems by adding a small amount of noise to
the evolution. A plot of the maximal Lyapunov exponent for elementary cellular
automata is reported in Fig.~\ref{figure:CALyapunov}.

\begin{figure}[t]
\centerline{\psfig{figure=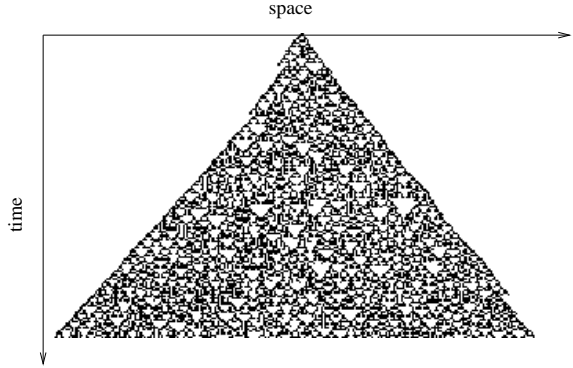,height=5cm,angle=270}}
\caption{The temporal plot of the damage spreading for rule 22.
\label{figure:Hamming22}
}
\end{figure}

The only source of randomness for deterministic cellular automata is in the
initial configuration. The counterpart of chaoticity is the dependence on the
initial condition. For chaotic automata a small change in the initial
configuration propagates to all the lattice. Measures of this kind are called
\textit{damage spreading} or \textit{replica symmetry breaking}. 

It is
convenient to introduce the difference field (damage)  $h_i^t$ between two
configurations $x_i^t$ and $y_i^t$
\begin{equation}
	h_i^t = x_i^t \XOR y_i^t
\end{equation}
and the Hamming distance $H(t)=1/n\sum_{i=1}^n h_i^t$. In
Fig.~\ref{figure:Hamming22} the spreading of the difference from a single site
for rule 22 is reported (see also Section~\ref{section:Damage}).

Finally, instead of studying the diverging rate of two trajectories, one can
measure the strength required to make all trajectories coalesce. Once again,
this is an indicator of the chaoticity of the automaton.

\subsection{Probabilistic Automata}

For probabilistic cellular automata the look-up table is replaced by a table
of transition probabilities that express the probability of obtaining
$\sigma_i^{t+1}=1$ once given the neighborhood configuration. For 
$k=1/2$ we have four transition probabilities
\begin{equation}
\begin{array}{rcl}
	\tau(0,0\go 1) &=& p_0; \\
	\tau(0,1\go 1) &=& p_1; \\
	\tau(1,0\go 1) &=& p_2; \\
	\tau(1,1\go 1) &=& p_3, 
\end{array} \label{probabilisticCA}
\end{equation}
with $\tau(a,b\go 0) = 1-\tau(a,b\go 1)$.

The evolution rule becomes a probabilistic Boolean function. This can be
written as a deterministic Boolean function of some random bits, thus allowing
the use of multi-site coding. In order to do that, one has to write the
deterministic Boolean functions that characterize the configurations in the
neighborhood with the same probability. For instance, let us suppose that the
$k=1/2$ automaton of equation~(\ref{probabilisticCA}) has to be simulated with
$p_0 = 0$, $p_1=p_2=p$ and $p_3=q$. The Boolean function that gives 1 only for
the configurations $(\sigma_{i-1}^t, \sigma_{i+1}^t) = (0,1)$ or $(1,0)$ is
\begin{equation}
\chi_i^t(1)=\sigma_{i-1}^t \XOR \sigma_{i+1}^t
\label{chi1}
\end{equation} 
and the function that
characterizes the configuration $(1,1)$ is 
\begin{equation}
  \chi_i^t(2)=\sigma_{i-1}^t \AND \sigma_{i+1}^t.
  \label{chi2}
\end{equation}

Let me introduce the truth function. The expression
$\tru{\mbox{\textit{expression}}}$ gives 1 if \textit{expression} is true and
zero otherwise (it is an extension of the delta function).
Given a neighborhood configuration $\{\sigma^t_{i-1}, \sigma^t_{i+1}\}$, 
$\sigma_i^{t+1}$ can be
1 with probability $p_i$. This can be done by extracting a random number $0\le
r<1$ and computing
\begin{equation}
	\sigma_i^{t+1} = \sum_i \tru{p_i > r}\chi_i.
	\label{probabilistic}
\end{equation}

Although the sum is not a bitwise operation, it can safely used here
since only one out of the $\chi_i$ can be one for a given configuration of the
neighborhood. The sum can be replaced with a \texttt{OR} or a
\texttt{XOR} operation.

Given a certain lattice configuration $\bold{a}=\bold\sigma^t$ at time $t$,
the local transition probabilities allow us to compute the transition
probability $\mcal{T}_{\bold{b}\bold{a}}$ from configuration $\bold{b}$ to
configuration $\bold{a}$ as
\begin{equation}
	\mcal{T}_{\bold{b}\bold{a}} = \prod_{i=1}^N \tau(a_{i-1}, 
	  a_{i+1} \go b_i). \label{DetailedBalance}
\end{equation}
 
\begin{figure}[t]
\centerline{\psfig{figure=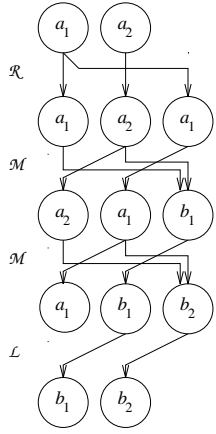,height=6cm}}
\caption{\label{figure:transfer} The application of the transfer matrix 
for the $k=1/2$, $d=1$, $N=2$ cellular automata.}
\end{figure}

The factorization of the matrix $\mcal{T}$ implies that it can be written
as a product of simpler transfer matrices $\mcal{M}$ that add only one site
to the configuration. Periodic boundary conditions require some attention: in
Fig.~\ref{figure:transfer} an example of decomposition
$\mcal{T}=\mcal{L}\mcal{M}^N\mcal{R}$ for the simplest case $N=2$
is shown. The lattice has been skewed for the ease of visualization.

This transition matrix defines a Markov process. 

The state of the system at a certain time $t$ is indicated by the probability
$x_{\bold{a}}^{(t)}$ of observing the configuration $\bold{a}$. 
Clearly, one has $\sum_{\bold{a}} x_{\bold{a}}^{(t)}=1$. 
For a deterministic automata only one component of $\bold{x}$ is
one, and all other are null. The time evolution of $\bold{x}$ is given by the
applications of the transfer matrix $\mcal{T}$. The conservation of
probability implies that $\sum_{\bold{a}} \mcal{T}_{\bold{b}\bold{a}} =1$. 
It is possible to
prove that the maximum eigenvalue $\lambda_1$ of
$\mcal{T}$ is one; the eigenvector corresponding to this eigenvalue is an
asymptotic state of the system.

If all configurations are connected by a chain of transition probabilities
(i.e.\ there is always the possibility of going from a state to another) than
the asymptotic state is unique.

The second eigenvalue $\lambda_2$ gives the correlation length $\xi$:
\begin{equation}
   \xi = -(\ln \lambda_2)^{-1}.
\end{equation}

In our system two different correlation length can be defined: one in the
space direction  and one in the time direction. 
The temporal correlation length gives
the characteristic time of convergence to the asymptotic state.

The usual equilibrium spin models, like the Ising or the Potts model, are
equivalent to a subclass of probabilistic cellular automata. In this case the
transition probability between two configurations $\bold{a}$ and $\bold{b}$ is
constrained by the detailed balance principle
\begin{equation}
\dfrac{\mcal{T}_{\bold{ba}}}{\mcal{T}_{\bold{ab}}} = 
  \exp\left(\beta\mcal{H}(\bold{b})-\beta\mcal{H}(\bold{a})\right),
\end{equation}
where $\mcal{H}(\bold{a})$ is the energy of configuration 
$\bold{a}$ and $\beta$ is the inverse of the temperature. 
One can invert the relation between equilibrium spin models and probabilistic
cellular automata~\cite{Georges} and reconstruct the Hamiltonian from the
transition probabilities. In general, a cellular automata is equivalent to an
equilibrium spin system if no transition
probabilities are zero or one. Clearly in this
case the space of configurations is connected.

These systems can undergo a phase transition, in which some observable
displays a non-analytic  behavior in correspondence of a well defined value of
a parameter. The phase transition can also be indicated by the non-ergodicity
of one phase. In other words in the phase where the ergodicity is broken the
asymptotic state is no more unique. An example is the ferromagnetic
transition; in the ferromagnetic phase there are two ensembles of states that
are not connected in the thermodynamic limit. In the language of the
transfer matrix this implies that there are two (or more) degenerate
eigenvalues equal to one. The corresponding eigenvectors can be chosen such
that one has positive components (i.e.\ probabilities) for the states
corresponding to one phase, and null components for the states corresponding
to the other phase, and vice versa.  Indeed, the degeneration of eigenvalues
correspond to the divergence of the (spatial and temporal) 
correlation lengths.

\begin{figure}[t]
\begin{center}
$\mcal{M} = \left(
  \begin{tabular}{llllllll}
    $ 1$ & $ 1 - p$ & $ 0$ & $ 0$ & $ 0$ & $ 0$ & $ 0$ & $ 0$ \\
    $ 0$ & $ 0$ & $ 1 - p$ & $ 1 - q$ & $ 0$ & $ 0$ & $ 0$ & $ 0$ \\
    $  0$ & $ 0$ & $ 0$ & $ 0$ & $ 1$ & $ 1 - p$ & $ 0$ & $ 0$ \\
    $ 0$ & $ 0$ & $ 0$ & $ 0$ & $ 0$ & $ 0$ & $ 1 - p$ & $ 1 - q$ \\
    $  0$ & $ p$ & $ 0$ & $ 0$ & $ 0$ & $ 0$ & $ 0$ & $ 0$ \\
    $ 0$ & $ 0$ & $ p$ & $ q$ & $ 0$ & $ 0$ & $ 0$ & $ 0$ \\
    $ 0$ & $ 0$ & $ 0$ & $ 0$ & $ 0$ & $ p$ & $ 0$ & $ 0$ \\
    $  0$ & $ 0$ & $ 0$ & $ 0$ & $ 0$ & $ 0$ & $ p$ & $ q$
  \end{tabular}
\right)$
\end{center}
\caption{\label{figure:TransferMatrix}Transfer Matrix $\mcal{M}$ for $N=2$.}
\end{figure}

It is well known that equilibrium models cannot show phase transitions at a
finite temperature (not zero nor infinite) in one
dimension. However, this is no more true if some transition probabilities is
zero or one violating the detailed balance Eq.~(\ref{DetailedBalance}).  In
effect, also the one dimensional Ising model  exhibits a phase transition at a
vanishing temperature, i.e.\ when some transition probabilities become
deterministic.  In particular one can study the phase transitions of models
with adsorbing states, that are configurations corresponding to attracting
points in the language of dynamical systems. An automata with adsorbing states
is like a mixture of deterministic and probabilistic rules. A system with
adsorbing states cannot be equivalent to an equilibrium model.   

Let me introduce here an explicit model in order to be more concrete:
the Domany-Kinzel model~\cite{DomanyKinzel}. This model is defined on the
lattice $k=1/2$ and the transition probabilities are
\begin{equation}
\begin{array}{lcr}
\tau(0,0\go 1)&=&0;\\
\tau(0,1\go 1)&=&p;\\
\tau(1,0\go 1)&=&p;\\
\tau(1,1\go 1)&=&q.
\end{array}
\label{DK}
\end{equation}

The configuration 0, in which all cells assume the value zero, is the adsorbing
state. Looking at the single site transfer matrix $\mcal{M}$ (reported in
Fig.~\ref{figure:TransferMatrix} for the simple case $N=2$), one can see that (for
$p, q<1$) every configuration has a finite probability of going into the
configuration 0, while one can never exit this state. In this simple
probability space a state (i.e.\ a vector) $\bold{v}=(v_0, \dots,v_7)$ that
corresponds to the single configuration  $\bold{a}$ is given by
$v_{\bold{a}}=\delta_{\bold{ab}}$. The configuration 0 corresponds to the state
given by the vector  $\bold{w}^{(1)}=(1,0,0,\dots)$.

Let me indicate with the symbols $\lambda_i, i=1,\dots, 8$ the eigenvalues of
$\mcal{M}$, with $||\lambda_1||>||\lambda_2||>\dots>||\lambda_8||$. The
corresponding eigenvectors are ${\bold{w}}^{(1)}$, $\dots$, 
${\bold{w}}^{(8)}$. Let us
suppose that they form a base in this space. The Markovian character of
$\mcal{M}$ implies that the maximum eigenvalue is 1, corresponding to the
eigenvector ${\bold{w}}^{(1)}$.

A generic vector $\bold{v}$ can be written as
\begin{equation}
\bold{v}=a_1\bold{w}^{(1)}+ a_2\bold{w}^{(2)}+\cdots
\end{equation}

\begin{figure}[t]
\centerline{\psfig{figure=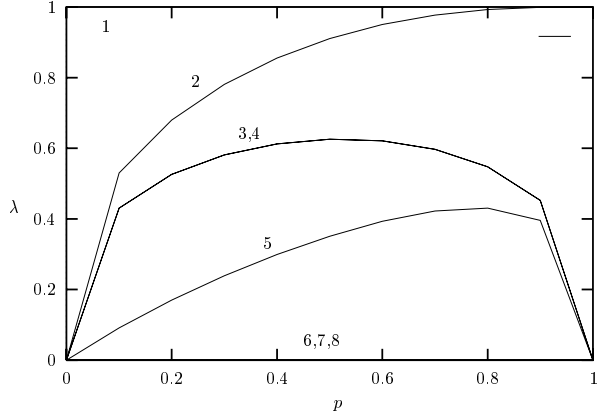,height=6cm}}
\caption{\label{figure:transfer1}Modulus of the eigenvectors of the $N=2$ 
 transfer matrix for $p=q$.}
\end{figure}

If we start from the vector $\bold{v}$ at time $t=0$ and we apply $T$ times
the transfer matrix $\mcal{T}$, in the limit of large $N$ this is
practically equivalent to the application of the matrix $\mcal{M}$ $NT$
times, and we get
\begin{equation}
\bold{v}(T,N) = \mcal{M}^{NT} \bold{v}(0,N) = 
   a_1\lambda_1^{NT}\bold{w}^{(1)}+ a_2\lambda_2^{NT}\bold{w}^{(2)}+\cdots
\end{equation}
and since all eigenvalues except the first one are in norm less that one, it
follows that the asymptotic state is given by the vector $w^{(1)}$ (i.e.\ by the
absorbing configuration 0).

\begin{figure}[t]
\centerline{\psfig{figure=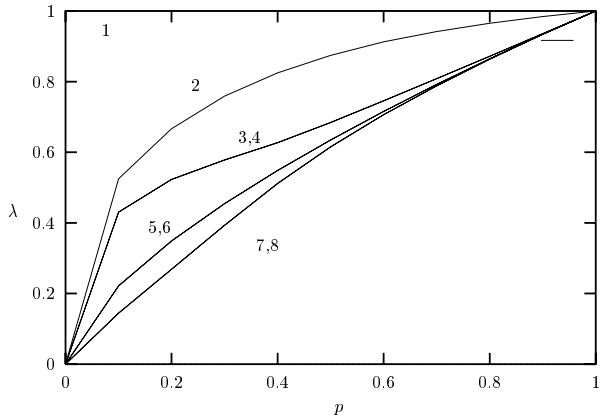,height=6cm}}
\caption{\label{figure:transfer2}Modulus of the eigenvectors of the $N=2$ 
transfer matrix for $q=0$.}
\end{figure}

This situation can change in the limit $N\go \infty$ and (after) $T\go\infty$
(the thermodynamic limit). In this limit some other eigenvalues can degenerate
with $\lambda_1$, and thus one or more configurations can survive forever. The
existence of such phase transitions can be inferred from the plot of the
modulus of the eigenvalues of $\mcal{M}$ for $N=2$. This is reported in
Fig.~\ref{figure:transfer1} for the case $p=q$. In this finite lattice the
degeneration with the eigenvalue $\lambda_1$ (which corresponds to the
configuration in which all sites take value 1) occurs for $p=1$. As discussed
in Section~\ref{section:Epidemics} in the thermodynamic limit the transition
occurs at $p=0.705\dots$. A rather different scenario exhibits for $q=0$. In
this case, as reported in Fig.~\ref{figure:transfer2}, all eigenvalues
degenerate with the first (for $p=0.81\dots$ in the thermodynamic limit) and
this implies a different dynamical behavior (see 
Section~\ref{section:Epidemics}).

\subsubsection{Critical Phenomena} 
\label{CriticalPhenomena}

The already cited phase transitions are a very interesting subject of study by
itself. We have seen that the correlation length $\xi$ diverges in the
vicinity of a phase transition. The correlation between two sites
of the lattice at distance $r$ is supposed to behave as $\exp(-r/\xi)$, the
divergence of $\xi$ implies very large correlations. Let us suppose that there
is a parameter $p$ that can be varied. In the vicinity of the critical value
$p_c$ of this parameter the correlation length diverges as $\xi\sim
(p-p_c)^{-\nu}$, where $\nu$ is in general non integer. Also other quantities
behaves algebraically near a critical point,
like for instance the magnetization $\rho$ which scales as
 $\rho\sim(p-p_c)^\beta$.

This power-law behavior implies that there is no characteristic scale of the
phenomena. Indeed, if we change the scale in which the parameter $p$ is
measured, the proportionality factor will change but the form of the law will
not. The pictorial way of explaining this phenomena is the following: suppose
we have some two-dimensional system near to a phase transition where one phase
is white and the other is black. There will be patches and clusters of all
sizes. If we look at this specimen by means of a TV camera, the finest details
will be averaged on. Now let us increase the distance from the TV camera and
the specimen (of infinite extension). Larger and larger details will be
averaged out. If the system has a characteristic scale $\xi$, the picture will
change qualitatively when the area monitored will be of order of the square of
this length or more. On the other hand, if we are unable to deduce the distance
of the TV camera from the surface by looking at the image, the system is
self-similar and this is a sign of a critical phenomena. Another sign is the
slow response to a stimulation: again, the distribution of the response times
follows a power-law.

The critical phenomena are very sensible to the parameter $p$, and a small
change in it will destroy this self-similarity. In spite of  their
non-robustness, critical phenomena are heavily studied because their
universality: since only large scale correlations are important, the details
of the rule do not change the exponents of the power laws. This implies that
the critical points of very different systems are very
similar.

Self-similar objects are very common in nature. One example is given by
clouds: it is almost impossible to  evaluate the distance from a cloud,
even when one is relatively near to it.  And also power laws are found in
many different fields, so are slow response times. It is impossible that
we always meet usual critical phenomena, which are so delicate. 		 	 It
has been proposed~\cite{SOC} that open systems (i.e.\ out of equilibrium
system) can auto-organize into a self-organized critical state, a state
that exhibits the characteristic of a critical point still being stable
against perturbations. It is indeed possible to develop simple models that
exhibit this feature, but it is not clear how ubiquitous this quality is. 
The Game of Life of Section~\ref{section:Life} is a very simple example of
a Self Organized Critical (SOC) model. 

\subsubsection{Diffusion} 
\label{section:Diffusion}

\begin{figure}[t]
\centerline{\psfig{figure=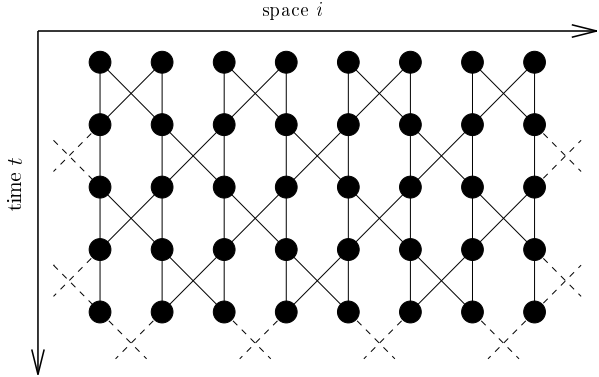,height=6cm}}
\caption{The lattice for the diffusion in $d=1$.}
\label{figure:twolattice}
\end{figure}

The implementation of diffusion in the context of cellular automata is not very
straightforward. One possibility is of course that of exchanging a certain
number of pairs of cells randomly chosen in the lattice. While this approach
clearly mixes the cells,~\footnote{For an application of this technique,
refer to the contribution by Boccara, this volume.} 
it is a serial method that cannot be applied in
parallel. A more sophisticated technique is that of dividing the lattice in
pieces (a one dimensional lattice can be divided in couples of two cells,
a two dimensional lattice in squares of four cells), and then rotate the cell in
a patch according with a certain probability. This is equivalent to
considering more complex lattices, but always with a local character, as shown
in Fig.~\ref{figure:twolattice} for $d=1$. 
The diffusion can be controlled by a parameter $D$ that gives the probability
of rotating the block. One has also to establish how many times the procedure
has to be repeated (changing the blocks). A detailed analysis of the effects of
this procedure can be found in Chopard and Droz (1989)~\cite{DrozChopard}. 

\section{Numerical techniques}
\label{section:Numerical}
 
In this section I shall review some techniques that I currently use to
investigate cellular automata. 

\subsection{Direct simulations}

The first approach is of course that of assigning each cell in the lattice to a
computer word, generally of type \texttt{integer}, and to write the evolution
rule using \texttt{if...} instructions. This generally produces very slow
programs and waste a large amount of memory (i.e.\ the size of the 
lattices are limited). 

Another possibility is to use look-up tables. In this case one adds the value
of the cells in the neighborhood multiplying them by an appropriate factor (or
packs them into the bits of a single word), and then uses this number as an
index in an array built before the simulation according with the rule.  

Alternatively, one can store the configurations in patches of 32
or 64 bits (the size of a computer word). This implies a
certain degree of gymnastic to make the patches to fit together at boundaries.

Again, since often one has to repeat the simulation starting from different
initial configurations, one can work in parallel on 32 or 64 replicas of the
same lattice, with the natural geometry of the lattice. When using
probabilistic cellular automata (i.e.\ random bits), one can use the same input
for all the bits, thus making this method ideal for the damage spreading
investigations (see Section~\ref{section:Damage}). 

Finally, one can exploit the natural parallelism of serial computers 
to perform in parallel simulations for the whole phase diagram, as described in
Bagnoli {\it et al.\ } (1997).~\cite{Bagnoli:Fragments}

In order to use the last three methods, all manipulations of the bits have
to be done using the bitwise operations \texttt{AND}, \texttt{OR},
\texttt{XOR}. It is possible to obtain a
Boolean expression of a rule
starting from the look-up table (canonical form), but this generally implies
many operations. There are methods to reduce the length of the Boolean
expressions.~\cite{Boolean,Bagnoli:Minimization}

\subsection{Mean Field}
\label{section:MeanField}

The mean field approach is better described using probabilistic cellular
automata. Given a lattice of size $L$, its state is defined by the probability
$x_{\bold{a}}$ of observing the configuration $\bold{a}$. 
If the correlation length $\xi$ is
less than $L$, two cell separated by a distance greater that $\xi$ are
practically independent. The system acts like a collection of subsystems each
of length $\xi$. Since $\xi$ is not known a priori, one assumes
a certain correlation length $l$ and thus a certain system size $L$, 
and computes the quantity of
interest. By comparing the values of these quantities
with increasing $L$ generally a clear scaling law appears, 
allowing to extrapolate the results to the case $L\go\infty$.

The very first step is to assume $l=1$. In this case the $l=1$ cluster
probabilities are $\pi_1(1)$ and $\pi_1(0)=1-\pi(1)$ for the normalization. The
$l=1$ clusters are obtained by $l=1+2k$ clusters via the transition
probabilities. 

For the $k=1/2$ Domany-Kinzel model we have
\begin{equation}
\begin{array}{rcl}
\pi'_1(1)&=&(\pi_2(0,1)+\pi_2(1,0)p + \pi_2(1,1) q. \\
\pi'_1(0)&= &\pi_2(0,0)+(\pi_2(0,1)+\pi_2(1,0)(1-p) + \pi_2(1,1)(1-q),
\end{array}
\end{equation} 
where $\pi=\pi^{(t)}$ and $\pi'=\pi^{(t+1)}$.

In order to close this hierarchy of equation, one factorizes the $l=2$
probabilities. If we call $\rho=\pi_1(1)$, we have $\pi_2(0,1) = \pi_2(1,0) =
\rho(1-\rho)$ and $\pi_2(1,1) = \rho^2$. The resulting map for the density
$\rho$ is
\begin{equation}
  \rho' = 2p\rho+(q-2p)\rho^2.
  \label{DK:MeanField}
\end{equation}
The fixed points of this map are $\rho=0$ and $\rho=2p/(2p-q)$. The stability
of these points is studied by following
 the effect of a small perturbation. We
have a change in stability (i.e.\ the phase transition) for $p=1/2$ regardless
of $q$.

The mean field approach can be considered as a bridge between cellular
automata and dynamical systems since it generally
reduces a spatially extended discrete
system to a set of coupled maps.

There are two ways of extending the above approximation. The first is still to
factorize the cluster probabilities at single site level but to consider more
time steps, the second is to factorize the probabilities in larger clusters.
The first approach applied for two time steps implies the factorization of
$\pi_3(a,b,c)=\pi_1(a)\pi_1(b)\pi_1(c)$ and the map is obtained by applying
for two time steps the transition probabilities to the $l=3$ clusters. The map
is still expressed as a polynomial of the density $\rho$. The
advantage of this method is that we still work with a scalar (the density),
but in the vicinity of a phase transition the convergence towards the
thermodynamic limit is very slow. 

The second approach, sometimes called \textit{local structure
approximation}~\cite{LocalStructure}, is a bit more complex. Let us start from
the generic $l$ cluster probabilities $\pi_l$. We generate the $l-1$ cluster
probabilities $\pi_{l-1}$ from $\pi_l$ by summing over one variable:
\begin{equation} 
	\pi_{l-1}(a_1, \dots,  a_{l-1}) = \sum_{a_l} \pi_{l}(a_1,
	\dots,  a_{l-1}, a_l). 
\end{equation} 
The $l+1$ cluster probabilities are
generated by using the following formula 
\begin{equation} 
	\pi_{l+1}(a_1,a_2,\dots,  a_l, a_{l+1})=
		\dfrac{\pi_{l}(a_1, \dots, a_l) \pi_{l}(a_2, \dots,
		a_{l+1})}{ \pi_{l-1}(a_2, \dots, a_{l})}. 
\end{equation} 
Finally, one is back
to the $l$ cluster probabilities by applying the  transition probabilities
\begin{equation} 
	\pi'(a_1, \dots, a_l)= \sum_{b_1,\dots,b_{l+1}}
	\prod_{i=1}^l\tau(b_i, b_{i+1}\go a_i). 
\end{equation}

This last approach has the disadvantage that the map lives in a
high-dimensional ($2^l$) space, but the results converges much better in
the whole phase diagram. 

This mean field technique can be considered an application of the
transfer matrix concept to the calculation of the the eigenvector corresponding
to the maximum eigenvalue (fundamental or ground state).

\subsection{Damage spreading and Hamming distance}
\label{section:Damage}

In continuous dynamical systems a very powerful indicator of chaoticity is the
Lyapunov exponent. The naive definition of the (maximum) Lyapunov exponent is
the diverging rate of two initially close trajectories, in the limit of
vanishing initial distance. 
 
This definition cannot be simply extended  to discrete systems, but we can
define some quantities that have a relation with chaoticity in dynamical
systems.

First of all we need a notion of distance in discrete space. The natural
definition is to count the fraction of corresponding cells that have different
value (I consider here only Boolean cellular automata).
 If we indicate with $\bold{x}$ and $\bold{y}$ the two configurations,
the distance $h$ between them (called the Hamming distance) is
\begin{equation}
	h=\dfrac{1}{n} \sum_{i=1}^n x_i \XOR y_i.
\end{equation}

It is possible to define an equivalent of a Lyapunov
exponent~\cite{Bagnoli:Lyapunov} (see Fig.~\ref{figure:CALyapunov}),
 but the natural application of the Hamming
distance is related to the damage spreading. 

For deterministic cellular automata the damage is represented by the Hamming
distance between two configurations, generally starting from a small number of
damaged sites, and the goal is to classify the automata according with the
average speed of the damage or other dynamical quantities.

The extension to probabilistic cellular automata is the following: what will happen
if we play again a rule with a different initial configuration but \textit{the
same realization of the noise}? If the asymptotic state changes, than the
evolution remembers the initial configuration, otherwise it is completely
determined by the noise. In the language of dynamical systems, this two
scenarios correspond to a breaking of replica (the two configurations)
symmetry. We can also define the difference field $h_i=x_i \XOR y_i$ with $h$
representing its density. The breaking of replica symmetry thus corresponds to 
a phase transition for $h$ from the adsorbing state $h=0$ to a \textit{chaotic}
state.

\section{Investigation Themes}

The theory described in the last section gives us the analytic tools and
constitutes an interesting  subject of study by itself. However, cellular
automata can be studied as phenomenological models that mimic the real world
from a mesoscopic point of view. The point of view of a physicist is again
that of looking for the simplest model still able to reproduce the qualitative
behavior of the original system. For this reason the models described in this
section will be very crude; if one wants a model that mimics the system as
good as possible, one can start from a simplified model and add all the
features needed. An advantage of cellular automata with respect to system of
differential or partial differential equations is the stability of dynamics.
Adding some feature or interactions never leads to structural instabilities.

This section is of course not exhaustive.

\subsection{Life}
\label{section:Life}

The Game of Life was introduced in the '70s by John Conway and then
popularized by Martin Gardner in the columns of Scientific
American~\cite{Life}. It is a two dimensional, Boolean,  outer totalistic,
deterministic cellular automata that in some sense resembles the evolution of
a bacterial population. The value 0 is associated to an empty or dead cell,
while the value 1 to a live cell.  The automaton is defined on a square
lattice and the neighborhood is formed by the nearest and next-to-nearest
neighbors. The evolution rule is symmetric in the values of the cells in the
outer neighborhood, i.e.\ it depends on their sum. The transition rules are
stated in a pictorial way as
\begin{itemize}
\item If a live cell is surrounded by less than two live cell, it will die by isolation.
\item If a live cell is surrounded by more that three live cells, it will die by overcrowding.
\item Otherwise, if a live cell is surrounded by two or three live cells, it will survive.
\item An empty cell surrounded by three live cells will become alive.
\item Otherwise an empty cell will stay empty.
\end{itemize}

\begin{figure}[t]
\centerline{\psfig{figure=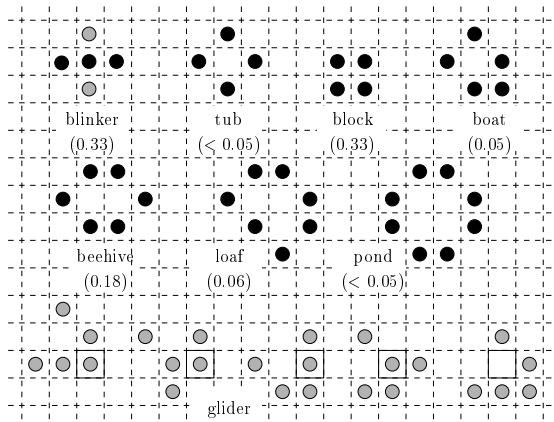,height=6cm}}
\label{figure:LifeAnimals}
\caption{The most common animals in Life. The figures represent the average
abundance of each animal.}
\end{figure}

The evolution of this rule is impressive if followed on the screen of a fast
computer (or a dedicated machine). Starting from a random configuration with
half cell alive, there is initially a rapid drop in the density $\rho^t$, 
followed by a
long phase of intense activity.~\footnote{An exhaustive analysis of the 
Game of Life can be found in Bagnoli {\it et al\ }
(1992).~\protect\cite{Bagnoli:Life}} 
 After some hundreds time steps (for lattices
of size order $200 \times 200$) there 
will emerge colonies of activity separated by nearly
empty patches. In these patches there are small stable or oscillating
configurations (the animals, see Fig.~\ref{figure:LifeAnimals}). Some of these
configurations can propagate in the lattice (the gliders). The activity zones
shrink very slowly, and sometimes a glider will inoculate the activity on a
quiescent zone. Finally, after hundreds time steps, the configuration will
settle in a short limit cycle, with density $\rho \simeq 0.028$. The scaling
 of the relaxation time with the size
of the lattice suggests that in an infinite lattice the activity will last
forever (see Fig.~\ref{figure:LifeTemporal}). The existence of a
non-vanishing asymptotic density has been confirmed by recent simulations
performed by Gibbs and Stauffer (1997)~\cite{Stauffer:life} 
on very large lattices (up to
$4 \times 10^{10}$ sites).

\begin{figure}[t]
\centerline{\psfig{figure=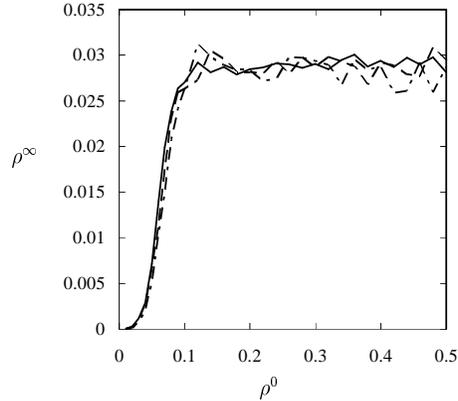,height=6cm}}
\caption{
\label{figure:LifeDensity}
The plot of the asymptotic density $\rho^\infty$ vs. the 
initial density $\rho^0$ in the Game
of Life. The curves represents different lattice sizes: $96\times 50$
 (broken line), $160\times100$ (dashed line), $320\times 200$ (full line).}
\end{figure}

There are several questions that have been addressed about the Game of Life. 
Most of them are mathematical, like the equivalence of Life to a universal
computer, the existence of \textit{gardens of heaven}, the feasibility of a
self-reproducing structure in Life, and so on (see 
Sigmund (1993)~\cite{Sigmund}).

From a statistical point of view, the main question concerns the dependence of
the asymptotic state from the density of the initial configuration (supposed
uncorrelated) and the response time of the quiescent state. 

\begin{figure}[t]
\centerline{\psfig{figure=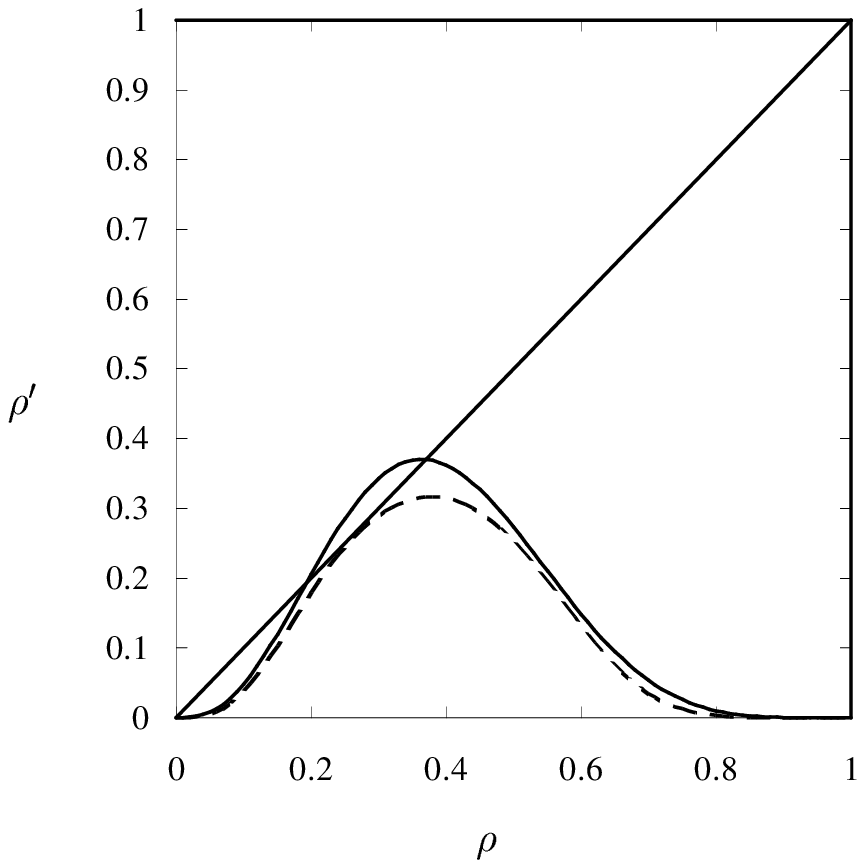,height=6cm}}
\caption{
\label{figure:LifeMF}
The mean field approximations for the Game of Life.}
\end{figure}

For the first question, in Fig.~\ref{figure:LifeDensity} is reported the plot
of the asymptotic density versus the initial density for a lattice square
lattice with $L=256$. One can see a transition of the value asymptotic
density with respect to the initial density for a value of the latter
around $0.05$. In my opinion this effect 
is due to the finite size of the lattice, since there is a slight
dependence on the system size. However, recent simulations performed by
Stauffer~\cite{Stauffer:privateLife} on very large lattices still exhibit
this effect.  

\begin{figure}[t]
\centerline{\psfig{figure=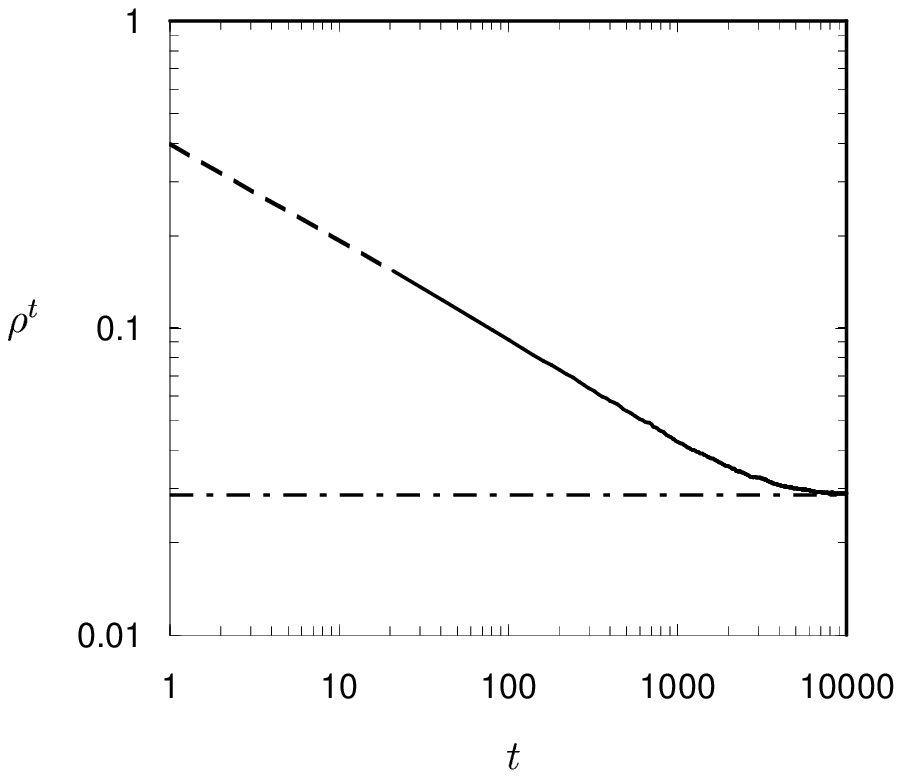,height=6cm}}
\caption{
\label{figure:LifeTemporal}
The temporal behavior of the density in the Game of Life.}
\end{figure}

The response time can be measured by letting Life to relax to the asymptotic
state, and then perturbing it in one cell or adding a glider. Recording the
time that it takes before relaxing again to a periodic orbit and plotting its
distribution, Bak {\it et al.}\ (1989)~\cite{LifeSOC} 
found a power law,
characteristic of self organized critical phenomena.
One can investigate also 
the dynamical properties of the Game of Life, for instance
the spreading of damage. Its distribution will probably follow a power law. 

Another possibility is to investigate the asymptotic behavior using mean field
techniques. The simplest approximations are unable to give even a rough
approximation of the asymptotic density $\rho \simeq 0.028$, so it is worth to
try more sophisticated approximations as the local structure one. 

\subsection{Epidemics, Forest Fires, Percolation}
\label{section:Epidemics}

We can call this class of processes \textit{contact processes}. They are
generally representable as probabilistic cellular automata. Let me give a
description in terms of an epidemic problem (non lethal)

In this models a cell of the lattice represents an individual (no empty cells),
and it can stay in one of three states: healthy and
susceptible (0), ill and infective (1), or immune (2). 

Let us discuss the case without immunization. This simple model can be
studied also in one spatial dimension (a line of individual) + time. 

Clearly the susceptible state is adsorbing. If there are no ill individual, the
population always stays in the state 0. We have to define a range of
interactions. Let us start with $k=1$, i.e.\ the state of one individual will
depend on its previous state and on that of its nearest neighbors. There is an
obvious left-right symmetry, so that the automata is outer totalistic.  The
probability of having $\sigma_i^{t+1} = 1$ will depend on two variables: the
previous state of the cell $\sigma_i^t$ and the sum of the states of
the neighboring cells. Let me use the totalistic characteristic functions
$\chi_i^t(j)$ that take the value one if the sum of the variables in the
neighborhood (here $\sigma_{i-1}^t$ and  $\sigma_{i+1}^t$) is $j$ and zero
otherwise (see Eq.~(\ref{chi1},~\ref{chi2})). Moreover, I shall denote 
$\sigma_i^t$ with $\sigma'$ and I shall neglect to indicate the spatial and
temporal indices. Then
\begin{equation}
\sigma' = f(\sigma, \chi(j)).
\end{equation}

\begin{figure}[t]
\centerline{
  \psfig{figure=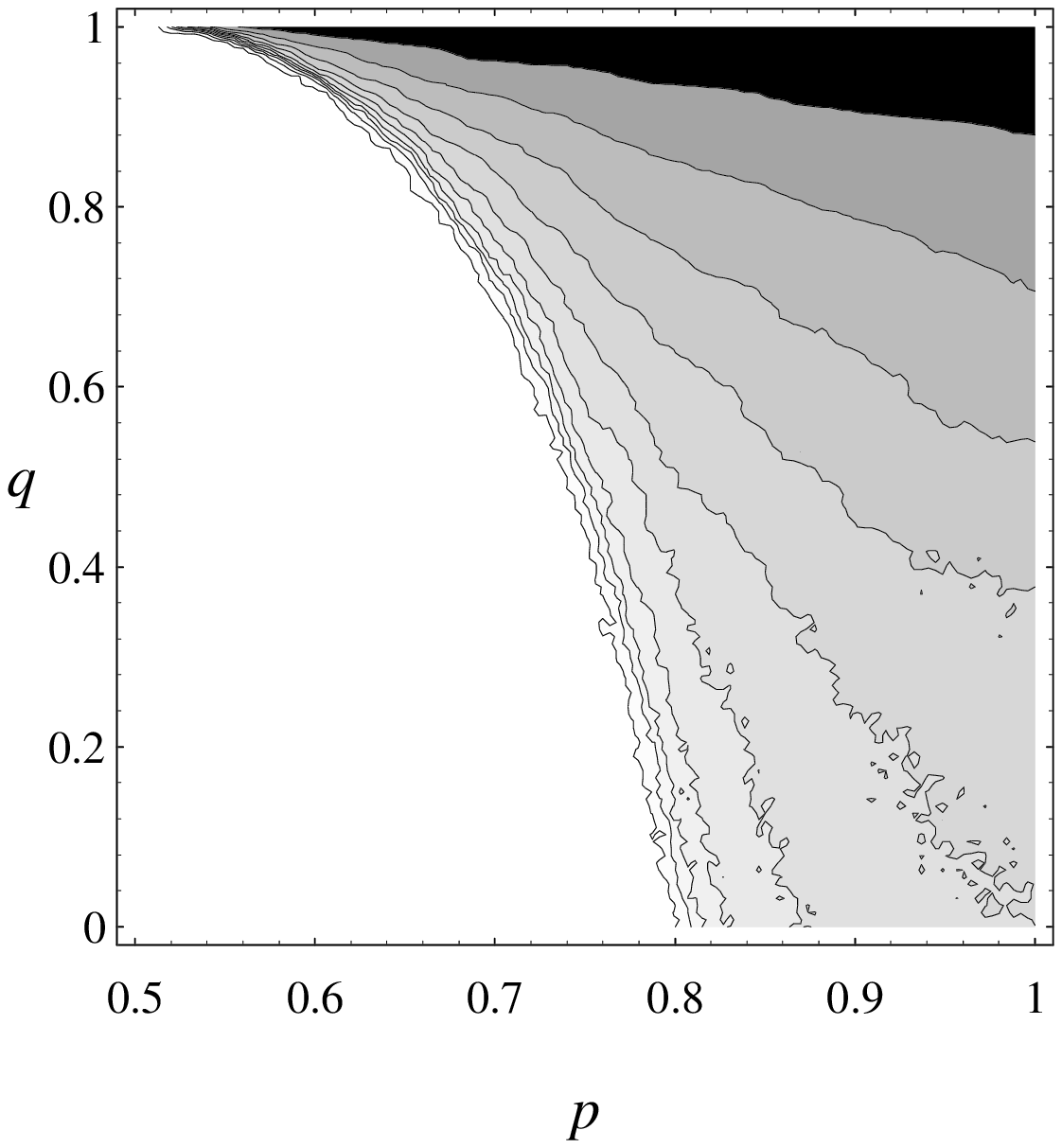,width=5cm,clip=}%
  \hspace{.6cm}%
  \psfig{figure=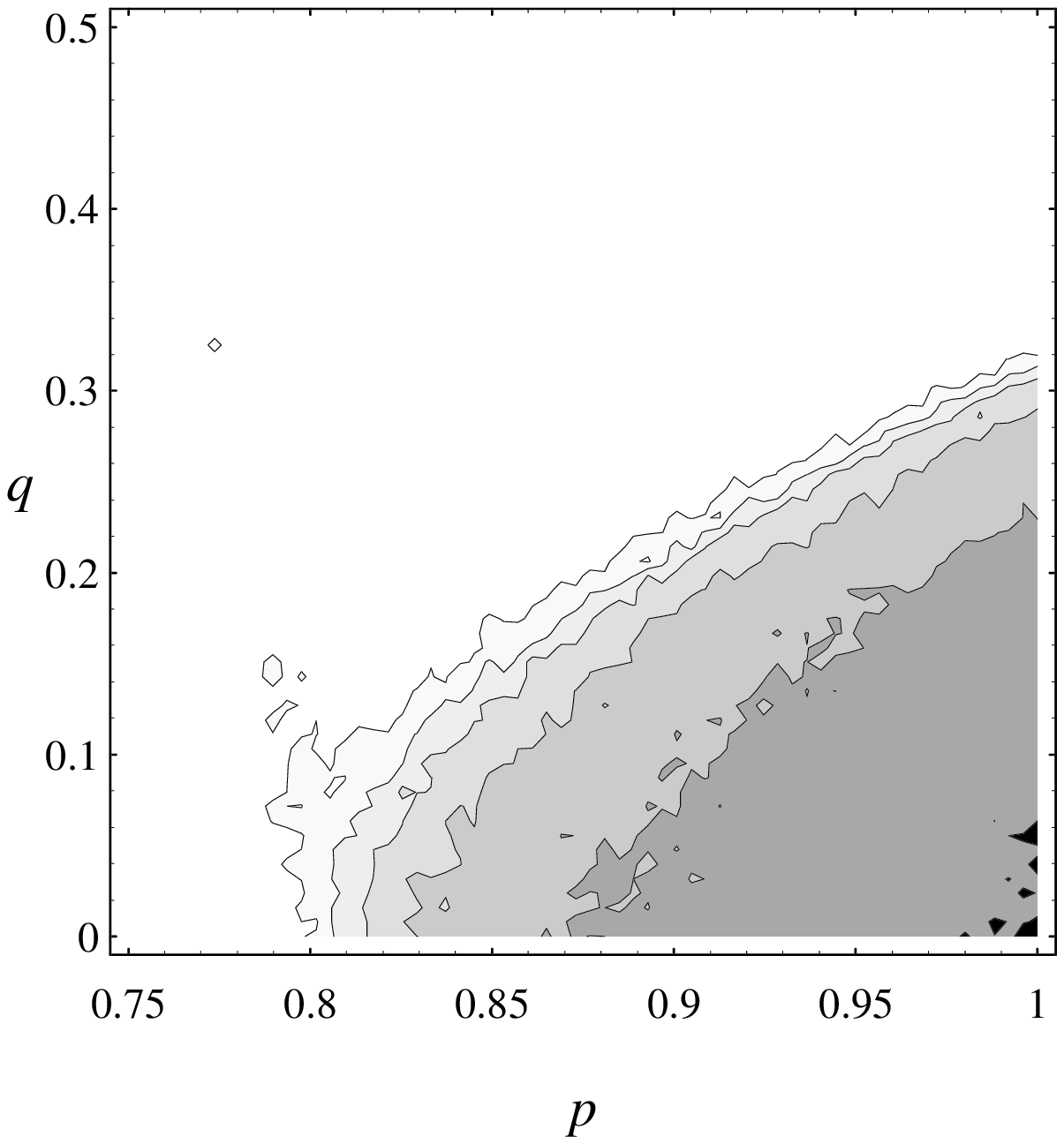,width=4.7cm,clip=}%
}
\caption{The phase diagram of the density (left) and damage (right)
for the Domany-Kinzel model.
 ($0\le p\le 1$, $0\le q\le 1$, $N=1000$, $T=1000$). The gray intensity is
 proportional to the value of the density, ranging from white to black in
 equally spaced intervals.}
\label{figure:DK}
\end{figure}

The function$f$ gives the probability that $\sigma'$ is ill (1) given its
present state and the number of ill neighbors. If $f$ does not depend on
$\sigma$ (i.e.\ the probability of contracting the infection is the same of
staying ill for a given number of ill neighbors) we have again the
Domany-Kinzel model Eq.~(\ref{DK}). 
The parameters of this model are here the probability of contracting the
infection or staying ill when an individual is surrounded by one ($p$) or two
($q$) sick individuals. The phase diagram of the model is reported in
Fig.~\ref{figure:DK}.

As one can see, there are two phases, one in which the epidemics last forever
(\textit{active} phase) and one in which the asymptotic state is only formed by
healthy individuals (\textit{quiescent} phase).

A feature of this model is the presence of a region in the phase diagram with
spreading of damage (\textit{active} phase). This zone corresponds to $p
\gg q$,
i.e.\ it is due to an interference effect between the
neighbors.~\footnote{See also Bagnoli (1996).~\protect\cite{Bagnoli:Damage}} 
This is
obviously not realistic for a generic infection, but it can have some
application to the spreading of a political idea. 

One can also study the case in which if an individual is surrounded by a
totality of ill individuals, it will certainly contract the infection
($q=1$). This
means that also the state in which all the cells have value one is adsorbing. 
In this case the transition is sharp (a first order transition). 

\begin{figure}[t]
\centerline{
  \psfig{figure=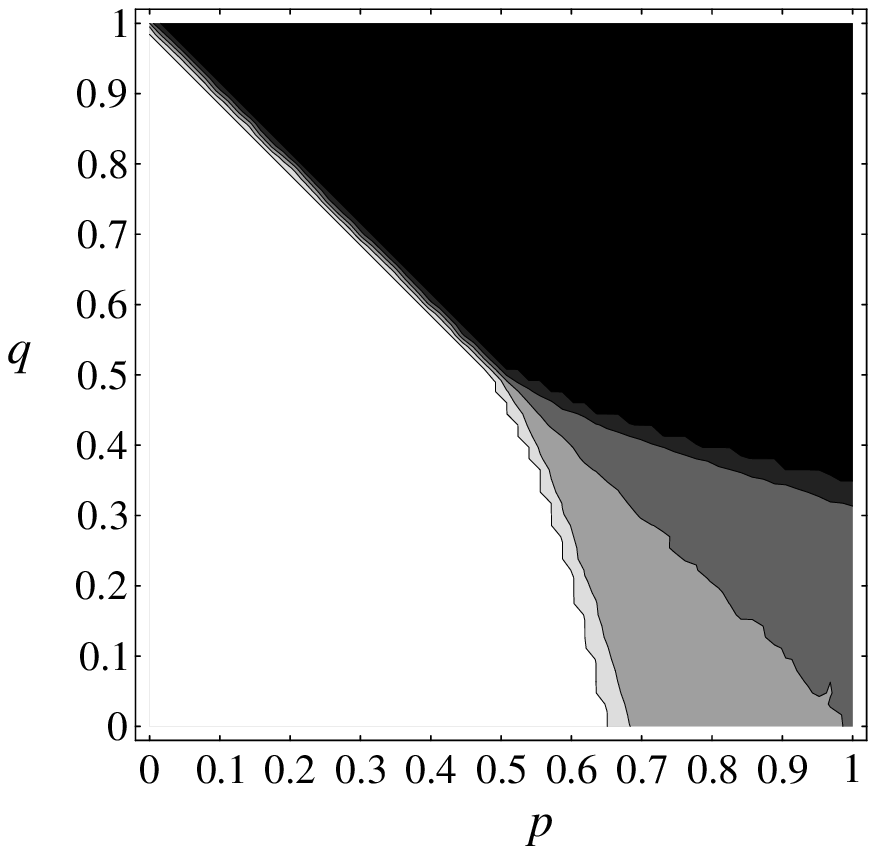,width=4.7cm,clip=}%
  \hspace{.6cm}%
  \psfig{figure=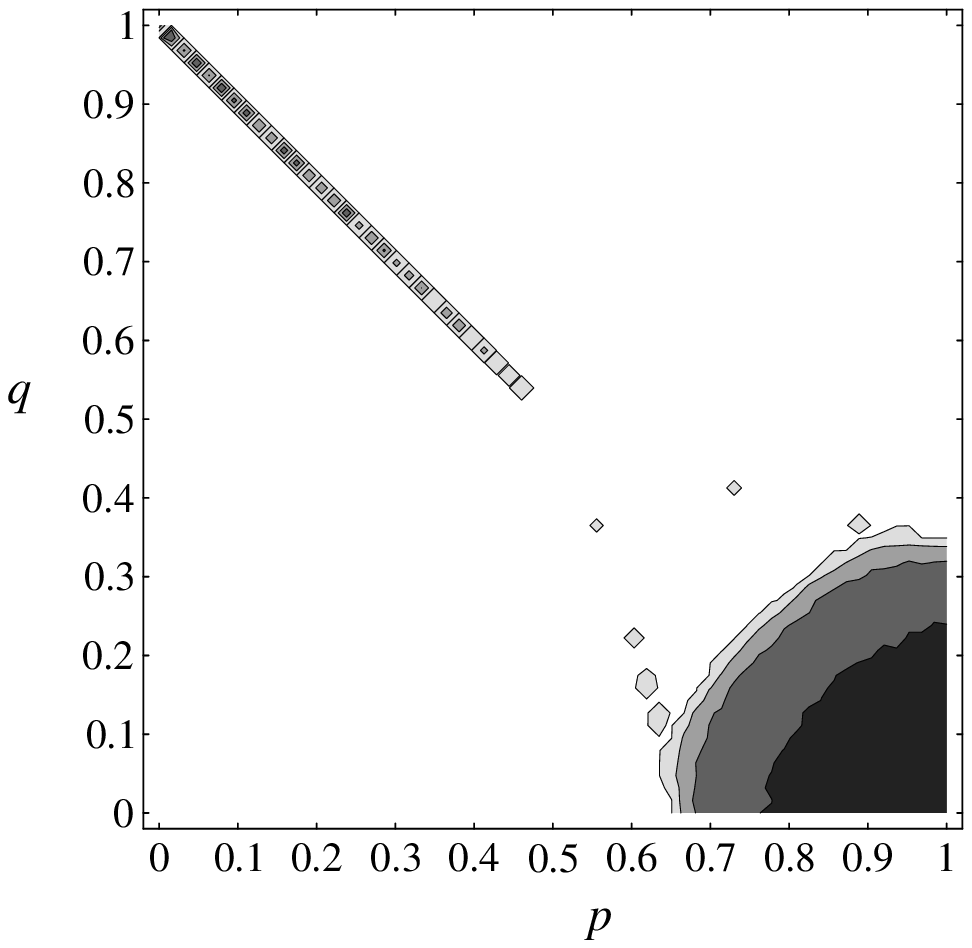,width=5.2cm,clip=}%
}
\caption{The phase diagram for the $k=1$ totalistic cellular automaton with two
adsorbing states.
\label{figure:Totalistic3}
}
\end{figure}

The phenomenology enriches if we allow the previous value of the cell $\sigma$
to enter the evolution function. Let us consider the a simplest case: a
totalistic function of the three cells. Let us call $p$, $q$ and $w$ the
probability of having $\sigma'=1$ if $\chi(1)$, $\chi(2)$ or $\chi(3)$ is one,
respectively. If we set $w=1$ we have two adsorbing states and two parameters,
so that the phase diagram is again two dimensional and easy to visualize (see
Fig.~\ref{figure:Totalistic3}). This model will be studied in detail in
the contribution  by  Bagnoli, Boccara and
Palmerini, this volume.  

We see that in this case we can have both first and second order phase
transitions. This observation could have some effective importance, in that a
first order phase transition exhibit hysteresis, so that if the system enters
one adsorbing state, it would cost a large change in the parameter to have it
switch to the other state. On the other hand, a second order phase transition
is a continuous one, with no hysteresis. 

The model can be extended to larger neighborhoods, immune
states, more dimensions and diffusion, always looking to the phase diagram and
to damage spreading. The mathematical tools that can be used are direct
simulations (with the fragment technique) and various mean
field approximations (once more, the local structure technique is quite
powerful).

\subsection{Ecosystems}
\label{section:Ecosystems}

I shall use the following definition of an ecosystem: an evolving set of
interacting individuals exposed to Darwinian selection. We  start from the
simplest model: let us consider an early ecosystem, composed by haploid
individuals (like bacteria). Each individual can sit on a cell of a
lattice in $d$ space dimensions.  Each
individuals is identified by its genetic information, represented as an
integer number $x$. The genotype can be read as a four symbol (bases) or
codon string.  One could also consider the genome composed by 
alleles of a set of genes. I shall
use here a binary coding (say, two alleles), because it is simpler to
describe.  All individuals have a genome of the same length $l$.  Thus, we
can interpret $x$ as a base-two number representing the genotype or as an
abstract index. The difference enters when we consider the mutations and
thus we have to introduce the concept of distance in the genetic
space.~\footnote{In effects, the  distance in the genetic space is defined in
terms of the number of mutations needed to connect (along the shortest
path) two individuals (or, more loosely,  two species).} In the first case
the genotype space is an hypercube with $2^l$ dimensions, in the second it
is an abstract space of arbitrary dimension, that for simplicity 
we can consider one dimensional. Finally, we have to introduce the phenotipic
distance, which is the difference in phenotipic traits between two individuals.
Given a genotype $x$, the phenotype is represented as $g(x)$. In order to
simplify the notations, I shall denote a function of the phenotype (say
$h(g(x))$ as $h[x]$. The phenotypic distance will be denoted as
$\delta(g(x),g(y)) = \delta[x,y]$.
I do not consider the influence of age, i.e.\ a genotype univocally determines
the phenotype. There could be more than one genotype that give origin to the
same phenotype (polymorphism).   

This automaton has a large number of states, one for each different genome
plus a state ($*$) for representing the empty cell.  Thus $*$ represents
an empty cell, 0 is the genome $(0,0,\dots,0)$, 1 is the genome
$(0,\dots,0,1)$ and so on. The evolution of the automaton is given by the
application of three rules: first of all the interactions among neighbors
are considered,
giving the probability of surviving, then we perform the diffusion step
 and finally
the reproduction phase.  I shall describe the one  (spatial)  dimensional
system, but it can be easily generalized.

\begin{description}

\item[survival:] An individual at site $i$ in the state $x_i^t \neq *$ 
and surrounded by $2k+1$ (including itself)  neighbors in the states
$\{x_{i-k}, \dots, x_{i+k}\}=\{x_i\}_k$ has a probability $\pi(x_i,
\{x_i\}_k)$ of surviving per unit of time. This probability is determined
by two factors: a fixed term $h[x_i]$ that represents its ability of
surviving in isolation, and an interaction term
1/(2k+1)$\sum_{j=i-k}^{i+k} J[x_i,x_j]$. Chearly, both the static fitness and
the interaction term depend on the phenotype of the individual.

 The fixed field $\bold{h}$ and the interaction matrix
$\mcal{J}$  define the chemistry of the world and  are fixed (at least in
the first version of the model). The idea is that the species $x$ with
$h[x]>0$ represent autonomous individuals that can survive in isolation
(say, after an inoculation into an empty substrate),  while those with
$h[x]<0$ represents predators or parasites that necessitate the presence
of some other individuals to survive. The distinction between autonomous
and non-autonomous species depends on the level of schematization. One
could consider an ecosystem with plants and animals, so that plants are 
autonomous and animals are not. Or one could consider the plants as a
substrate (not influenced by animals) and then herbivores are autonomous and
carnivores are not. The interaction matrix specifies the necessary inputs for
non autonomous species. 

We can define the fitness $H$~\footnote{There are several definition of
fitness, the one given here can be connected to the growth rate  $A$	of a
genetically pure population by $A = \exp(H)$; see also
Section~\protect\ref{Eco:MeanField}}  
of the species $x_i$ in the environment $\{x_i\}_k$ as
 \begin{equation}
  H(x_i, \{x_i\}_k) = h[x_i] + \dfrac{1}{2k+1}\sum_{j=i-k}^{i+k} J[x_i,x_j].
  \label{fitnessH}
\end{equation}

The survival probability $\pi(H)$ is given by some sigma-shaped function
of  the fitness, such as 
\begin{equation}
  \pi(H) = \dfrac{e^{\beta H}}{1+e^{\beta H}}=
     \dfrac{1}{2} + \dfrac{1}{2} \tanh(\beta H),
\end{equation}
where $\beta$ is a parameter that can be useful to modulate the effectiveness
of selection. 

The survival phase is thus expressed as:
\begin{itemize}
\item If $x_i \ne *$ then
\begin{equation}
\begin{array}{rcll}
x'_i&=&x_i\qquad &\mbox{with probability $\pi(H(x_i, \{x_i\}_j))$} \\
x'_i&=& *        &\mbox{otherwise}
\end{array}
\end{equation} 
\item Else
\begin{equation}
	x'_i = x_i = *\hspace{5cm}
\end{equation}
\end{itemize}

\item[diffusion:] The diffusion is given by the application of the procedure
described in Section~\ref{section:Diffusion}. The disadvantage of this approach
is that we cannot introduce nor an intelligent diffusion (like escaping from
predators or chasing for preys), nor different diffusion rates for different
species. An alternative could be the introduction of diffusion on a pair basis:
two cells can exchange their identity according with a given set of
rules.~\footnote{There are several ways of dividing a lattice into couples
of neighboring cells. If one want to update them in parallel, and if the
updating rule depends on the state of neighboring cells, one cannot update
a couple at the same time of a cell in the neighborhood. Since one has to
update a sublattice after another, there is a slightly asymmetry that
vanishes in the limit of a small diffusion probability iterated over
several sublattices.}

For the moment we can use the following rule:
\begin{itemize}
\item Divide the lattice in neighboring pairs (in $d=1$ and no dependence
on the neighbors  there are two ways of
doing that)
\item Exchange the values of the cells with
probability $D$
\end{itemize}

Clearly the probability $D$ could depend on the value of the cells and on that
of  neighbors.

\item[reproduction:] The reproduction phase can be implemented as a rule for
empty cells: they choose one of the neighbors at random and copy its identity
with some errors, determined by a mutational probability $\omega$. 

Thus
\begin{itemize}
\item If the cell has value $*$ then 
\begin{itemize}
\item choose one of the neighbors;
\item copy its state;
\item for each bits in the genome 
replace the bit with its opposite with probability $\omega$;
\end{itemize}
\item Else do nothing.
\end{itemize}
\end{description}

We have now to describe the structure of the interacting matrix $\mcal{J}$.
I shall deal with a very smooth correspondence between phenotype and genotype:
two similar genotypes have also similar phenotype. 
With this assumptions $x$ represents a strain rather than a species.
This hypothesis implies the
smoothness of $\mcal{J}$.

First of all I introduce the intraspecies competition. Since the individuals
with similar genomes are the ones that share the largest quantity of
resources, then the competition is stronger the nearer the genotypes.
This implies that $J[x,x]<0$. 

The rest of $\mcal{J}$ is arbitrary. For a classification in terms of usual
ecological interrelations, one has to consider together $J[x,y]$ and $J[y,x]$.
One can have four cases:

\begin{center}
\begin{tabular}{ccl}
$J[x,y]<0$ & $J[y,x]<0$ & competition \\
$J[x,y]>0$ & $J[y,x]<0$ & predation or parasitism \\
$J[x,y]<0$ & $J[y,x]>0$ & predation or parasitism \\
$J[x,y]>0$ & $J[y,x]>0$ & cooperation 
\end{tabular}
\end{center}

One has two choices for the geometry of the genetic space. 

\begin{description}

\item[hypercubic distance:] The genetic distance between $x$ and $y$ is given 
by the number of
bits that are different, i.e. $d(x,y) = ||x \XOR y||$, where the norm
$||x||$ of a Boolean vector $x$ is defined as $||x||=(1/N)\sum_{i=1}^N x_i$. 

\item[linear distance:] The genetic  space is 
arranged on a line (excluding the 0 and with periodic
boundary conditions)
rather than on a hypercube. This arrangement  is simpler but less
biological;~\footnote{An instance of a similar (sub-)space in real
organisms is  given by a
repeated gene (say a tRNA gene): a fraction of its 
 copies can mutate, linearly varying the
fitness of the individual with the ``chemical composition''
of the gene.~\protect\cite{BagnoliLio} This degenerate case has been
widely studied (see for instance Alves and Fontanari
(1996)~\protect\cite{Alves}). 
The linear space is equivalent to the hypercubic space if the phenotype $g(x)$
does not depend on the disposition of bits in $x$ (i.e.\ it depends only on the
number of ones -- a totalistic function):
one should introduce in this case
the
multiplicity of a degenerate state, which can be approximated to a
Gaussian, 
but if one works in the neighborhood of 
its maximum (the most common chemical composition) the
multiplicity factors are nearly constants. 
Another example is given by the level of catalytic
activity of a protein.  A linear space has also been used for modeling the
evolution of RNA viruses on HeLa cultures.~\protect\cite{Kessler}}

\end{description}

\subsection {Mean field approximation}\label{Eco:MeanField}

Since the model is rather complex, it is better to start\footnote{We also
stop in this lecture with this approximation.} with the 
mean field approximation, disregarding the spatial structure. 
This approximation becomes exact in the limit of large diffusion or when the
coupling extends on all the lattice.~\footnote{Since the original model is
still in development, I shall not study the exact mean field
version of the above description, but a simplified one.}

Let $n(x)$ be the number of organisms with genetic code $x$, and $n_*$ the
number of empty sites. if $N$ is the total number of cells, we have 
\begin{eqnarray*}
n_*+\sum_x n(x) &=&N; \\
m =\dfrac{1}{N}\sum_x n(x) &=&1-\dfrac{n_*}{N};
\end{eqnarray*}
considering the sums extended to all ''not-empty'' genetic codes,
and indicating with $m$ the fraction of non-empty sites (i.e.\ the population
size). 

We shall label with a tilde the quantities after the
survival step, with a prime after the reproduction step. The evolution of the
system  will be ruled
by the following equations:

\begin{eqnarray}
\tilde n(x) &=&\pi(x,n)n(x);  \label{model} \\
n^{\prime }(x) &=&\tilde n(x)+\frac{\tilde n_*}N\sum_yW(x,y)\tilde n(y). 
\nonumber
\end{eqnarray}
The matrix $W(x,y)$ is the probability of mutating from genotype $y$ to
 $x$. For a given codon mutation probability $\mu$, $W(x,y)$ is given by

\begin{description}

\item[hypercubic distance:] 
\begin{equation}
W(x,y) = \mu^{d(x,y)} (1-\mu) ^{l-d(x,y)};
\end{equation}

\item[linear distance:] 
\begin{equation}
\begin{array} {rcll}
 W(x,y) &= & \mu\qquad& \mbox{if $|x-y|=1$}; \\
 W(x,x) &=& 1-2\mu; & \\
 W(x,y) &=&0&\mbox{otherwise.}
\end{array}
\end{equation}
\end{description}

In any case the mutations represent a diffusion propcess in genic space, thus
the matrix $W$ conserves the total population, and 
\begin{equation}
	\sum_x W(x,y) = \sum_y W(x,y) = 1. \label{normW} 
\end{equation}

Summing over the genomes $x$, we can rewrite Eq.~(\ref{model}) for $m$ as
\begin{eqnarray*}
\widetilde{m} &=&\dfrac{1}{N}\sum_x\tilde n(x)=\dfrac{1}{N}\sum_x\pi(x,n)n(x) \\
m^{\prime } &=&\dfrac{1}{N}\sum_xn^{\prime }(x)=\widetilde{m}+\dfrac{\tilde n_*}{%
N^2}\sum_{xy}W(x,y)\tilde n(y).
\end{eqnarray*}

Introducing the probability
distribution of occupied sites $p(x)=\frac{n(x)}{mN}$ ($\sum_xp(x)=1$), and the
average fitness $\overline{\pi}$ as
\[
	\overline{\pi}\equiv \frac 1{mN}\sum_x\pi(x,n)n(x)=\sum_x\pi(x)p(x),
\]
and thus 
\[ 
	\widetilde{m}=m\overline{\pi}; \qquad  \frac{\tilde n_*}N=1-m\overline{\pi}.
\]
Using the property (\ref{normW}) we obtain
\begin{equation}
		m^{\prime } = m\overline{\pi}(2-m\overline{\pi}).\label{ecologistic}
\end{equation}

The stationary condition ($m^{\prime }=m$) gives
\begin{eqnarray*}
 1&=&\overline{\pi}(2-m\overline{\pi});\\
m &=&\frac{2\overline{\pi}-1}{\overline{\pi}^2}.
\label{mequation}
\end{eqnarray*}

The normalized evolution equation for $p(x)$ is thus
\begin{equation}
p^{\prime }(x)= \dfrac{\pi(x,p,m)p(x)+(1-m\overline{\pi})
	\sum_yW(x,y)\pi(x,p,m)p(y)}%
 {\overline{\pi}(2-m\overline{\pi})};  \label{peq}
\end{equation}
where the usual fitness function, Eq.(\ref{fitnessH}), 
has to be written in term of $p(x)$
\begin{eqnarray*}
H(x,p,m) &=&h[x]+m\sum_y J[x,y]p(y)
\end{eqnarray*}

Notice that Eq.(\ref{ecologistic}) corresponds to the usual logistic equation 
for population dynamics if we keep the average fitness
$\overline{\pi}$ constant.

\subsection{Speciation due to the competition among strains}

One of the feature of the model  is the appearance of the species intended as a
cluster of strains connected by mutations.~\cite{BagnoliBezzi}

One can consider the following analogy with a Turing mechanism for chemical
pattern formation. The  main
ingredients are an autocatalytic reaction process (reproduction) with slow
diffusion (mutations) coupled with the emission of a short-lived,
fast-diffusing inhibitor (competition). In this way a local high
concentration
of autocatalytic reactants inhibits the growth in its neighborhood, acting
as a
local negative interaction. 

In genetic space, the local coupling is given by the competition among
genetically kin individuals. For instance, assuming  a certain distribution
of some resource (such as some essential metabolic component for a
bacterial   population), then the more genetically similar two individuals
are,
the wider the fraction of shared resources is. The effects of 
competition on strain $x$ by strain $y$  are modeled by a term proportional
to
the relative abundance of the latter, 
$p(y)$, modulated by a function that decreases with the
phenotypic distance between $x$ and $y$. Another example of this kind of
competition  can be found in  the immune response in mammals. Since the
immune
response has a certain degree of specificity, a viral strain $x$ can suffer
from the response triggered by strain $y$ if they are sufficiently near in
an
appropriate phenotypic subspace.  Again, one can think that this effective
competition can be modeled by a term,  
proportional to the relative abundance of the strain that
originated the response, 
which decreases with the phenotypic distance. 

Let us start with a one dimensional ``chemical'' model of cells that
reproduce
asexually  and slowly diffuse (in real space), $p=p(x,t)$ being their
relative
abundance at position $x$ and at time $t$. These cells constitutively emit
a
short-lived, fast-diffusing mitosys inhibitor $q=q(x,t)$. This inhibitor
may be simply identified with some waste or with 
the consumption of a local resource (say
oxygen).  
The diffusion of the inhibitor is modeled as
\begin{equation}
	\dfrac{\partial q}{\partial t} = k_0 p +
		 D \dfrac{\partial ^2q}{\partial x^2} -k_1 q,\label{q}
\end{equation}
where $k_0$, $k_1$ and $D$ are the production, annihilation
and diffusion rates of $q$.

The evolution of the distribution $p$ is given by
\begin{equation} 	
	\dfrac{\partial p}{\partial t} = \left(A(x, t) -\overline{A}(t)\right)p + 
		\mu \dfrac{\partial ^2p}
		{\partial x^2},\label{p} 
\end{equation}
\begin{equation} 
	\overline{A}(t) = \int A(y,t)p(y,t){\rm d} y.\label{A}
\end{equation}
The growth rate $A$ can be expressed in terms of the fitness $H$ as
\begin{equation} 
	A(x,t) = \exp\left(H(x,t)\right).\label{H}
\end{equation}
Due to the form of equation~(\ref{p}), 
at each time step we have
\begin{equation}
\sum_{x}p(x,t)=1.
\label{norm}
\end{equation}

The
diffusion rate of $q$, $D$, is assumed to be much larger than $\mu$. The 
growth rate $A$,  can be decomposed in two
factors, $A(x,t)  = A_0(x) A_1(q(x,t))$,
where $A_0$ gives the reproductive rate in absence of $q$, so $A_1(0) = 1$.
In
presence of a large concentration of the inhibitor $q$ the reproduction
stops,
so $A_1(\infty)=0$. A possible choice is 
\[ 
A(x,t) = \exp(H_0(x) -  q(x,t)).
\] 
For instance, $H_0(x)$ could model the sources of food or, for algae
culture, the distribution of light. 

Since we assumed a strong separation in time scales, we look for a
stationary distribution $\tilde q(x,t)$ of the inhibitor
(Eq.~(\ref{q})) by keeping $p$ fixed. This  is given 
 by a convolution of the distribution $p$:
\[
	\tilde q (x,t) = J\int  \exp\left(-\dfrac{|x-y|}{R}\right) p(y,t) {\rm
	d}y, 
\] 
where $J$ and $R$ depend on the parameters $k_0$, $k_1$, $D$. In the
following
we shall use $J$ and $R$ as control parameters, disregarding their origin.

We can generalize this scenario to non-linear diffusion processes of the
inhibitor by using 
the reaction-diffusion equation Eq.~(\ref{p}), with the fitness $H$ 
and the kernel $K$ given by
\begin{equation}
	H(x,t) = 	H_0(x) - J\int K\left(
			\dfrac{x-y}{R}\right) p(y,t) {\rm d}y \label{kernel}
\end{equation}
\begin{equation}
	K(r) = \exp\left(-\dfrac{|r|^\alpha}{\alpha}\right),
	\label{K}
\end{equation}
i.e. a symmetric decreasing function of $r$ with 
$K(0)=1$. The parameters $J$ and $\alpha$ control
the intensity of the competition and 
the steepness of the interaction, respectively.    

Let us consider the correspondence with the genetic space:  the quantity
$x$ now
identifies a genome, the diffusion rate $\mu$ is given by mutations, and
the
inhibitor $q$ (which is no more a real substance)
represents the competition among phenotypically related strains. The
effects of competition are much faster than the genetic drift  (mutations),
so
that the previous hypotheses are valid. While the competition  interaction
kernel
$K(r)$ is not given by a diffusion process, its general form should be
similar
to that of Eq.~(\ref{K}): a decreasing  function of the phenotypic distance
between
two strains.  
We shall refer to 
the $p$-independent contribution to the fitness, $H_0[x]$,
as the static fitness landscape. 

Our model is thus defined by Eqs.~(\ref{p}--\ref{K}). 
We are interested in its asymptotic behavior in the limit
$\mu\rightarrow 0$. Actually, the mutation mechanism is needed only to
define
the genetic distance and to allow population of an eventual niche.
The results should not change qualitatively if one includes more realistic
mutation mechanisms.

Let us first examine the behavior of Eq.~(\ref{p}) in absence of
competition ($J=0$) for a smooth static landscape and 
 a vanishing mutation rate. This
corresponds to the Eigen model in one dimension: 
since it does not exhibit any phase transition, the
asymptotic distribution is unique. 
The asymptotic distribution is given by one delta function peaked
around the global maximum of the static landscape, or more delta functions 
(coexistence) if
the global maxima are degenerate. 
The effect of a small mutation rate is simply that of broadening
the distribution from a delta peak to a bell-shaped
curve~\cite{Bagnoli:ecal}. 

\begin{figure}[t]
\centerline{\psfig{figure=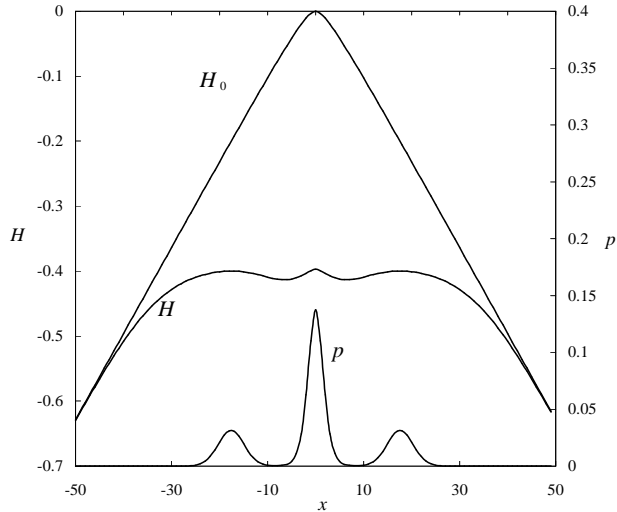,width=8cm}}
\caption{\label{fitness} 
Static fitness $H_0$, effective fitness 
$H$,  and asymptotic distribution $p$ 
numerically computed for the following values of
parameters: $\alpha=2$, $\mu=0.01$, $H_0=1.0$, 
$b=0.04$, $J=0.6$, $R=10$ and $r=3$.}
\end{figure}

While the degeneracy of maxima of the  static fitness
landscape is a very particular condition, 
we shall show in the following that in
presence of competition this is a generic case. 
For illustration, we report 
in Fig.~\ref{fitness} 
the numerical computation of the asymptotic behavior of the 
model
for a possible evolutive
scenario that leads to the coexistence of three species. We have chosen a
smooth static fitness $H_0$ (see Eq.~(\ref{H0}))
and a Gaussian ($\alpha=2$) competition 
kernel, using the genetic distance in place of the phenotypic one.
The effective fitness  $H$ is almost 
degenerate (here $\mu>0$ and the competition effect extends on
the neighborhood of
the maxima), and this leads to the coexistence.

In the following we shall assume again 
that the phenotypic space is the same of the
genotypic space.

\subsubsection{Evolution near a maximum}
\label{section:evolution_near_maximum}

We need the expression of $\bold{p}$ if a given static fitness $H(x)$ or a
static growth rate $A(x)$
has a smooth, isolated maximum for $x=0$ ({\it smooth maximum}
approximation).
Let us assume that 
\begin{equation}
A(x)\simeq A_{0}(1-ax^{2}), 
\label{pot}
\end{equation}
where $A_0 = A(0)$.

The generic evolution equation (master equation) for the
probability distribution is 
\begin{equation}
\alpha (t)p(x,t+1)=\left(1+\mu \frac{\delta^2}{\delta x^2}\right) 
	A\bigl(x,\bold{p}(t)\bigl)p(x,t);
	\label{alphap}
\end{equation}
where the discrete second derivative $\delta^2/\delta x^2$ is defined as
\[
\frac{\delta f(x)}{\delta x ^2} = f(x+1) + f(x-1) - 2 f(x),
\]
and $\alpha (t)$ maintains the normalization of $\bold{p}(t)$.
In the following we shall mix freely the 
continuous and discrete formulations of the problem.
 
Summing over $x$ in Eq.~(\ref{alphap}) and using the 
normalization condition, Eq.~(\ref{norm}),
we have: 
\begin{equation}
\alpha =\sum_{x}A(x,\bold{p})p(x)=\overline{A}. 
\end{equation}
The normalization factor $\alpha $ thus corresponds to the average 
fitness. The quantities $A$ and $\alpha $ are defined up to an 
arbitrary constant.

If $A$ is sufficiently smooth (including the dependence on
$\bold{p}$), one can
rewrite Eq.~(\ref{alphap}) in the asymptotic limit, using a continuous
approximation for $x$ as 
\begin{equation}
\alpha p=Ap+\mu \frac{\partial ^{2}}{\partial x^{2}}(Ap),
\label{alphap_cont}
\end{equation}
Where we have neglected to indicate
the genotype index $x$ and the explicit dependence on $\bold{p}$.
Eq.~(\ref{alphap_cont}) has the form of 
 a nonlinear diffusion-reaction equation. Since we want to 
 investigate the phenomenon of species
formation, we look for an asymptotic distribution $\bold{p}$ formed
by a 
superposition of
several non-overlapping bell-shaped curves, where the term non-overlapping 
means almost uncoupled by mutations. Let us number these curves using 
the index $i$, and denote each of them as $p_i(x)$, with
$p(x)=\sum_i p_i(x)$. Each $p_i(x)$ is centered around 
$\overline{x}_i$ and its weight is $\int p_i(x)dx=\gamma _i$, with 
$\sum_i\gamma _i=1$. We further assume that each $p_i(x)$ obeys
 the same asymptotic condition, Eq.~(\ref{alphap_cont}) (this is 
 a sufficient but not necessary condition). Defining 
\begin{equation}
\overline{A}_i=\frac{1}{\gamma _i}\int A(x)p_i(x)dx=\alpha,
\end{equation}
we see that in a stable ecosystem all quasi-species have the same average
fitness.

Substituting $q=Ap$ in Eq.~(\ref{alphap_cont}) we have (neglecting to
indicate
the genotype index $x$, and using primes to denote differentiation with
respect
to it): 
\[
\dfrac{\alpha }{A}q=q+\mu q''. 
\]
Looking for $q=\exp (w)$, %
\[
\dfrac{\alpha }{A}=1+\mu ({w'}^{2}+w''), 
\]
 and approximating $A^{-1}=A_0^{-1}\left( 1+ax^{2}\right) $, we have 
\begin{equation}
\dfrac{\alpha }{A_{0}}(1+ax^{2})=1+\mu ({w'}^{2}+w''). \label{alphaA0}
\end{equation}
A possible solution is 
\[
w(x)=-\dfrac{x^{2}}{2\sigma ^{2}}. 
\]
Substituting into Eq.~(\ref{alphaA0}) we finally get 
\begin{equation}
\dfrac{\alpha }{A_{0}}=\dfrac{2+a\mu - \sqrt{4a\mu +a^{2}\mu ^{2}}}{2}.
 		\label{smooth}
\end{equation}
Since $\alpha =\overline{A}$, $\alpha /A_{0}$ is less than one we have
 chosen the minus sign. In the limit $a\mu \rightarrow 0$ (small mutation
 rate
and smooth maximum), we have 
\[
\dfrac{\alpha }{A_{0}}\simeq 1-\sqrt{a\mu } 
\]
and 
\begin{equation}
\sigma ^{2}\simeq \sqrt{\dfrac{\mu }{a}}. \label{sigma}
\end{equation}

The asymptotic solution is 
\[
p(x)=\gamma \dfrac{1+ax^{2}}{\sqrt{2\pi }\sigma (1+a\sigma ^{2})}\exp \left(
-\dfrac{x^{2}}{2\sigma ^{2}}\right),
\]
so that $\int p(x)dx=\gamma $. The solution is a bell-shaped curve, its
width $\sigma$ being determined by the combined effects
of the curvature $a$ of maximum and the mutation rate $\mu$.. 
In the next section, we shall apply these results to a quasi-species $i$.
In this case one should substitute $p \rightarrow p_i$, 
$\gamma \rightarrow \gamma_i$ and $x \rightarrow x-\overline{x}_{i}$. 

For completeness, we study also the case of a {\it sharp maximum},  
for which  $A(x)$ varies considerably with $x$. In this case 
the growth rate of less fit strains has a 
large contribution from the mutations of fittest strains, 
while the reverse flow is negligible, thus
\[
p(x-1)A(x-1) \gg p(x)A(x) \gg p(x+1)A(x+1) 
\]
neglecting last term, and substituting  $q(x)=A(x)p(x)$ in
Eq.~(\ref{alphap})
we get: 
\begin{align}
	\dfrac{\alpha}{A_0}  = 1-2\mu &\qquad \mbox{for $x=0$}\label{map0}\\
 	q(x) =\dfrac{\mu}{\left(\alpha A(x) 
 	-1+2\mu\right)} q(x-1)&\qquad \mbox{for
 	$x>0$} \label{map}
\end{align}

Near $x=0$, combining Eq.~(\ref{map0}), Eq.~(\ref{map})and
Eq.~(\ref{pot})), 
we have
\[
q(x) =\dfrac{\mu}{(1-2\mu)a x^{2}} q(x-1). 
\]

In this approximation the solution is
\[
q(x) = \left(\dfrac{\mu}{1-2\mu a}\right)^x \dfrac{1}{(x!)^2},
\]
and 
\[
y(x) = A(x)q(x) \simeq \dfrac{1}{A_0}(1+a x^2)\left(\dfrac{\mu A_0}{\alpha
a}\right)^x \dfrac{1%
}{x!^2}. 
\]

\begin{figure}[t]
\centerline{\psfig{figure=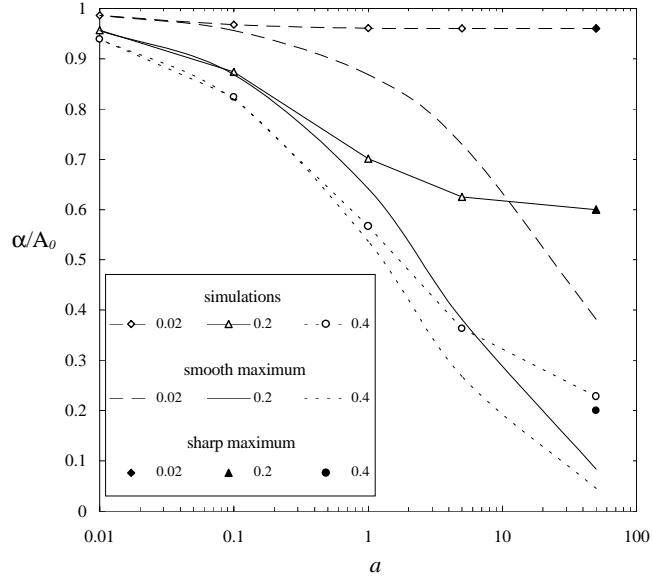,width=9cm}}
\caption{ Average fitness $\alpha/A_{0}$ versus the coefficient $a$,
of the fitness function, Eq.~(\ref{pot}),  for some values of the mutation
rate $\mu$. Legend: {\it numerical resolution} corresponds to the numerical
solution of 
Eq.~(\ref{alphap}),  
{\it smooth maximum} refers to  Eq.~(\ref{smooth}) and {\it sharp maximum}
to Eq.~(\ref{map0})
\label{alpha}
}
\end{figure}

One can  check the validity of these approximations by solving  numerically 
Eq.~(\ref{alphap}); 
the comparisons are shown 
in Fig.~\ref{alpha}.
One can check that the {\it smooth maximum} approximation agrees with the
numerics for
for small values of $a$, 
 when $A(x)$ varies slowly with $x$, while the {\it
sharp maximum} approximation  agrees with the  numerical results for large
 values of  $a$, when small variations of $x$ correspond to large
variations of $A(x)$. 

\subsubsection{Speciation}

I shall now derive the conditions for the coexistence of multiple species.
Let us assume that the asymptotic distribution is formed by $L$ delta
peaks $p_k$, $k=0, \dots, L-1$,
 for a vanishing mutation rate (or $L$ non-overlapping bell shaped
curves for a small mutation rate) centered at $y_k$.
The weight of each quasi species is $\gamma_k$, i.e.
\[
	 \int p_k(x) dx = \gamma_k, \qquad\sum_{k=0}^{L-1} \gamma_k = 1.
\]
The quasi-species are ordered such as  $\gamma_0 \ge
\gamma_1,  \dots, \ge \gamma_{L-1}$. 

 The evolution equations for the $p_k$ are 
($\mu \rightarrow 0$) 
\[
	\dfrac{\partial p_k}{\partial t} = (A(y_k) - \overline A) p_k,
\]
where $A(x) = \exp\left(H(x)\right)$ and
\[
	H(x) = 
		H_0(x) - J\sum_{j=0}^{L-1} K\left(\dfrac{x - y_j}{R}\right) \gamma_j.
\]

The stability condition of the asymptotic distribution is 
$(A(y_k) - \overline A) p_k = 0$, i.e. either
 $A(y_k) = \overline A = \text{const}$
(degeneracy of maxima) or $p_k=0$ (all other points). In other terms one
can
say that in a stable environment the fitness of all individuals is the
same,
independently on the species. 

The position $y_k$ and the weight $\gamma_k$ of the quasi-species
are given by $A(y_k) = \overline A = \text{const}$ and 
$\left.{\partial A(x)}/{\partial x}\right|_{y_k} = 0$, or, in terms of the
fitness $H$, by
\[
	H_0(y_k) - J \sum_{j=0}^{L-1} K\left(\dfrac{y_k-y_j}{R}\right)
		 \gamma_j = \text{const}
\]
\[
	H'_0(y_k)  - \dfrac{J}{R}\sum_{j=0}^{L-1} K'\left(\dfrac{y_k-y_j}{R}\right)
	 \gamma_j = 0
\]

Let us compute the phase boundary for coexistence of three species for two
kinds of kernels: the exponential (diffusion) one ($\alpha=1$)
and a Gaussian one ($\alpha=2$). 

I assume that the static fitness $H_0(x)$ is a symmetric
linear decreasing function
except in the vicinity of $x=0$, where it has a quadratic maximum:
\begin{equation}
	H_0(x) = b\left(
		1-\dfrac{|x|}{r} - \dfrac{1}{1+|x|/r}
	\right)\label{H0}
\end{equation}
so that close to $x=0$ one has 
$H_0(x) \simeq -b x^2/r^2$ and for $x\rightarrow \infty$,
$H_0(x) \simeq b(1-|x|/r)$. Numerical simulations show that the results
are qualitatively independent on the exact form of the static fitness,
providing that it is a smooth decreasing function. 

Due to the symmetries of the problem, we have one quasi-species at $x=0$
and
two symmetric quasi-species at $x=\pm y$. Neglecting the mutual influence
of
the two marginal quasi-species, and considering that $H'_0(0) = K'(0)=0$, 
$K'(y/R) = -K'(-y/r)$, $K(0)=J$ 
and that the three-species threshold is given by $\gamma_0=1$ and
$\gamma_1=0$,
 we have 
\[
	\tilde{b}\left(1-\dfrac{\tilde{y}}{\tilde{r}}\right) 
				- K(\tilde{y}) = -1,  
\]
\[
	\dfrac{\tilde{b}}{\tilde{r}} + K'(\tilde{y}) = 0.
\]
where $\tilde{y}=y/R$, $\tilde{r} = r/R$ and $\tilde{b} = b/J$. 
I introduce the parameter $G=\tilde{r}/\tilde{b}=
(J/R)/(b/r)$, that is the ratio of two
quantities, one related to the strength of inter-species interactions
($J/R$) and the other to intra-species ones ($b/r$). 
In the following I shall drop the tildes for convenience.
Thus
\[
 r - z - G \exp\left(-\dfrac{z^\alpha}{\alpha}\right) = -G,
\]
\[
 G z^{\alpha-1}\exp\left(-\dfrac{z^\alpha}{\alpha}\right) = 1,
\]

For $\alpha=1$ we have the coexistence condition
\[
 \ln(G) = r -1 + G.
\]
The only parameters that satisfy these equations are $G=1$ and $r=0$,
i.e.\ a
 flat landscape ($b=0$) with infinite range interaction ($R=\infty$). 
Since the coexistence region reduces to a single point,
it is suggested that $\alpha=1$ is a marginal case.
Thus for less steep potentials, such as power law decrease, 
the coexistence condition is supposed not to be
fulfilled.  

For $\alpha=2$ the coexistence condition is given by
\[
	z^2 -(G+r)z + 1 = 0,
\]
\[
	Gz\exp\left(-\dfrac{z^2}{2}\right) = 1.
\]
One can solve numerically this system and obtain the boundary 
$G_c(r)$ for the coexistence. In the limit $r \rightarrow 0$ (static
fitness
almost flat) one has 
\begin{equation}
	G_c(r) \simeq G_c(0) - r \label{Gc}
\end{equation}
with $G_c(0) = 2.216\dots$. 
Thus for $G>G_c(r)$ we have coexistence of three or more quasi-species,
while 
for $G<G_c(r)$ only the fittest one survives.
\begin{figure}[t]
\centerline{\psfig{figure=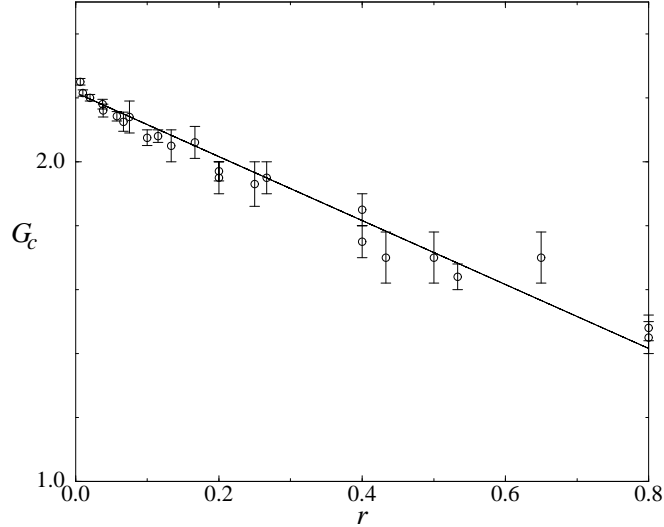,width=9cm,angle=270}}
\caption{Three-species coexistence boundary 
$G_c$ for $\alpha=2$. The continuous 
line represents the analytical 
approximation, Eq.~(\ref{Gc}), the circles are obtained from
numerical simulations. The error bars represent the maximum error.}
\end{figure}

I have solved numerically  Eqs.~(\ref{p}--\ref{K}) for several
different values of the parameter $G$,
considering  a discrete genetic space, with $N$ points, and a simple
Euler
algorithm. The results, presented in Fig.~2, are not 
strongly affected by the integration step.
The error bars are due to the
discreteness of the changing parameter $G$. 
The boundary of the multi-species phase is well approximated by
Eq.~(\ref{Gc});
in particular, I have checked that this 
 boundary does not depends on the mutation rate
$\mu$, at least for $\mu < 0.1$, which can be considered
a very high mutation rate for
real organisms. The most important effect of $\mu$ is the broadening of
quasi-species curves, which can eventually merge.

\subsection{Final considerations}

Possible measures on this abstract ecosystem model 
concern the origin of  complexity (for instance
considered equivalent to the entropy) in biological systems.
One expects that steady ecosystems reach an asymptotic value
of the complexity that maximizes the explotation of energy, while perturbed
ecosystems exhibit lower complexity. 
Another interesting topic regards the behavior of complexity in relation with
the spatial extension of a system.

\end{document}